\documentclass[]{aastex631}

\usepackage{amsmath}
\begin{document}

\title{Abundances and Transients from Neutron Star-White Dwarf Mergers}

\author[0000-0002-9604-7908]{M. Alexander Kaltenborn}
\affiliation{Department of Physics and Astronomy, 
  the George Washington University,
  Washington, DC 20052, USA}
\affiliation{Theoretical Division, Los Alamos National Laboratory, 
  Los Alamos, NM 87545, USA}
\affiliation{Center for Theoretical Astrophysics, Los Alamos National Laboratory, Los Alamos, NM, 87545, USA}

\author[0000-0003-2624-0056]{Chris L. Fryer}
\affiliation{Department of Physics and Astronomy, 
  the George Washington University, 
  Washington, DC 20052, USA}
\affiliation{Center for Theoretical Astrophysics, Los Alamos National Laboratory, Los Alamos, NM, 87545, USA}
\affiliation{Computer, Computational, and Statistical Sciences Division, Los Alamos National Laboratory, Los Alamos, NM, 87545, USA}
\affiliation{The University of Arizona, Tucson, AZ 85721, USA}
\affiliation{Department of Physics and Astronomy, The University of New Mexico, Albuquerque, NM 87131, USA}

\author[0000-0003-3265-4079]{Ryan~T. Wollaeger}
\affiliation{Center for Theoretical Astrophysics, Los Alamos National Laboratory, Los Alamos, NM, 87545, USA}
\affiliation{Computer, Computational, and Statistical Sciences Division, Los Alamos National Laboratory, Los Alamos, NM, 87545, USA}
\affiliation{Department of Physics and Astronomy, Louisiana State University, Baton Rouge, LA 70803, USA}

\author[0000-0002-1658-7681]{Krzysztof Belczynski}
\affiliation{Nicolaus Copernicus Astronomical Center, The Polish Academy of Sciences, 18 Bartycka Street, 00-716 Warsaw, Poland}

\author[0000-0002-5412-3618]{Wesley Even}
 \affiliation{Theoretical Division, Los Alamos National Laboratory, 
  Los Alamos, NM 87545, USA}
  \affiliation{Center for Theoretical Astrophysics, Los Alamos National Laboratory, Los Alamos, NM, 87545, USA}

\author[0000-0003-1443-593X]{Chryssa Kouveliotou}
\affiliation{Department of Physics and Astronomy, 
  the George Washington University, 
  Washington, DC 20052, USA}



\begin{abstract}
We systematically investigate the mergers of neutron star-white dwarf binaries from beginning to end, with focus on the properties of the inflows and outflows in accretion disks and their electromagnetic emissions. Using population synthesis models, we determine a subset of these binaries in which the white dwarf companion undergoes unstable mass transfer and complete tidal disruption, forming a large accretion disk around the neutron star. The material evolves according to an advection-dominated accretion disk model with nuclear burning, neutrino-emissions, and disk-surface wind ejection. The extreme dynamics of the entire process has proven difficult for analytic analysis, and thus currently the properties are poorly understood. The outflows from the mergers are iron- and nickel-rich, giving rise to optical and infrared emissions powered from the decay of the radioactive iron-type isotopes, calculated via the {\it SuperNu} light-curve code. We find these systems capable of powering bright, yet short-lived, optical transients with the potential to power gamma-ray bursts.
\end{abstract}


%% and \citet commands to identify citations. The citations are

\section{Introduction} \label{sec:intro}
Mergers of compact stellar objects have been the center of astrophysical interest since the detection of the first Gamma-ray Bursts (GRBs). GRBs were first detected in 1967 with the Vela satellites while monitoring for nuclear detonations \citep{Klebesadel1973}. This led to a flurry of research into the origin of these energetic bursts. It was subsequently determined that the progenitors were extragalactic, but the mechanisms responsible for GRBs have been a challenge to pin down, largely due to the fact that light-curves produced by GRBs are extremely diverse. The distribution of GRB durations indicate a strong bimodality between short- and long-duration bursts\citep{Kouveliotou1993}. While the most widely accepted mechanism for the long-duration GRBs is the collapsar model \citep{MacFadyen1999}, the favored mechanism powering the short-duration bursts has been the merger of compact binary systems \citep{Kochanek1993,Bloom1999,Popham1999,Fryer1999b}. These mergers, at the extremes of known physical systems, offer a better understanding of the characteristics of our Universe. Not only are these mergers strong candidates for the source of many GRBs, but they are also the suspected sources for many other of the most energetic cosmic transients, such as kilonovae (KNe), the newly detected gravitational wave (GW) signals, and nurseries for the heaviest elements found in the Universe. The first detections of GWs associated with binary black hole (BH) and binary neutron star (NS) mergers have caused a spike of renewed interest into binary mergers \citep{GW150914,GW170817,GW170817EMC}. While binary BH-BH, NS-NS, and BH-NS have been extensively studied for their electromagnetic transient counterparts and GW signals \citep{GW190521,Narayan1992,Metzger2010,Nakar2020}, binary mergers of white dwarfs (WDs) with NS companions have not been explored in the same depth. 

There have been a multitude of prior studies of binary mergers involving BHs and WDs \citep{Fryer1999a}. These mergers are largely governed by the exact physics that determines the merger of NS-WD binaries, but the mass ratios between the two objects are significantly different than that of the less massive binaries. The mass ratio of the binary is the dominant characteristic dictating the early-time dynamics of the mass transfer, tidal disruption of the WD, and initial conditions of the accretion disk. Since NS-WD binaries occupy a different section of the mass-ratio band, it is worth studying specifically the consequences of these mergers. The interest into NS-WD mergers has grown in recent years, in particular regarding the characteristics of these binaries as well as the accretion disks and their properties. These works have inspired our systematic study across the binary composition range, in particular work by Metzger and Margalit \citep{Metzger2010,Margalit2016}.

NS-WD binary systems are uniquely positioned to explain specific subsets of detectable signals. The BH-NS and double NS mergers have been extensively studied; these are commonly associated with GRBs that are less than two seconds \citep{Berger2014}. These bursts are driven by the remaining material from the merger event that is collected into a disk and accreted onto the merger remnant. While WDs merging with a NS or BH also produce accretion disks, the disks formed are unique in structure. These disks are expected to contain a larger fraction of the initial WD mass. This results in disks with more angular momentum, larger radii, and slower accretion rates than those formed from the more massive and compact binaries. The rate of BH-WD mergers is much lower than that of NS-WD mergers \citep{Fryer_1999}, so any emissions will be dominated by the latter. It is likely that the distinct characteristics of these progenitors will be associated with a distinct subset of observable transients, potentially describing a portion of the diverse variety of optical transients, GRB light curves, and GW signals.

Not only are the properties of the formed accretion disk and the potential transients specific to NS-WD mergers, but also these mergers are expected to be much more common than their more energetic cousins. The number of NS-WD binaries that have been detected are significant, and population synthesis calculations point to there being a large number of these binaries within our own Galaxy ($2.2\times 10^6$ binaries in the Milky Way) \citep{Nelemans2001a,Toonen2018}, with merger rates calculated to be between $8-500$ Myr$^{-1}$. Merger rates have more recently been calculated to be from $1-2 \times 10^{-4}$ yr$^{-1}$ up to $8 \times 10^{-6} - 5 \times 10^{-4}$ yr$^{-1}$, which is in agreement with the previous population synthesis rate of $1.4 \times 10^{-4}$ yr$^{-1}$ \citep{Nelemans2001a} and the binary pulsar measurement rate of $2.6 \times 10^{-4} $yr$^{-1}$ \citep{Bobrick2017}. In the near future, when these more frequent binary mergers do occur within our Galaxy, not only will we have a chance at observing them in the electromagnetic regime, but we will also have the opportunity to capture them using GW detectors like the LISA constellation \citep{babak2021lisa}. Thompson et al.(2009) suggest a rate of 25 Myr$^{-1}$, resulting in at least one merger detected with LISA per year. Because of their populations and rates of coalescence NS-WD systems deserve further studies as they can also help potentially identify new families of transients.

In this paper, we explore NS-WD binaries through their life cycles, specifically focusing on their end state. In Section~\ref{sec:nswdproperties}, we briefly explore the initial creation of these compact binaries, collect the populations of different NS-WD binaries, determine the orbital separation at which the merger begins, and discuss the nature of the unstable mass transfer in order to select our binary configurations for further study. In Section~\ref{sec:accmodel}, we introduce the analytic approximations of the accretion disks that will be used to study the properties of the mergers. Section~\ref{sec:numresults} contains the results from the model outlined herein, with a focus on the resulting nuclear abundances in the disk and the wind. In Section~\ref{sec:observables}, using these yields and abundances we review the possible transients one might expect to find resulting from NS-WD binaries. 

\section{Properties of NS-WD Progenitors}\label{sec:nswdproperties}
The NS-WD binaries begin their lives as binaries composed of massive stars. These stars undergo a cataclysmic event of some sort, collapsing into a degenerate companion. This can occur via core-collapsed supernovae (CCSNe) forming a NS or the ejection of the outer envelopes forming a WD. The secondary companion is formed later through other processes, e.g., from mass accretion from the companion. How these binaries come to be is a complex subject of study on its own and is beyond the scope of this paper. However, a brief understanding of population synthesis models and simulations is important for extracting expectations of number of binaries, composition of the binaries, and their merger rates.

\subsection{\label{subsec:popsynth}NS-WD Population Synthesis}
The properties of compact-object binaries have been studied with observational and theoretical methods. Here, the populations are obtained using the population synthesis code StarTrack \citep{Belczynski2008}. StarTrack is able to evolve isolated single stars and binaries for a wide range of initial conditions. The input physics incorporates up-to-date knowledge of processes that govern stellar evolution, while the most uncertain aspects of the evolution are parameterized to allow for systematic error analysis. The populations of the compact binaries are gathered by evolving the initial stellar components using analytic formulas, taking into account orbit circularization due to tidal interaction, magnetic braking, gravitational wave radiation, mass exchange via Roche-lobe overflow, common envelopes, and empirical parameterizations of wind-mass loss. 

\begin{figure}[h]
\centering
\includegraphics[width=\textwidth]{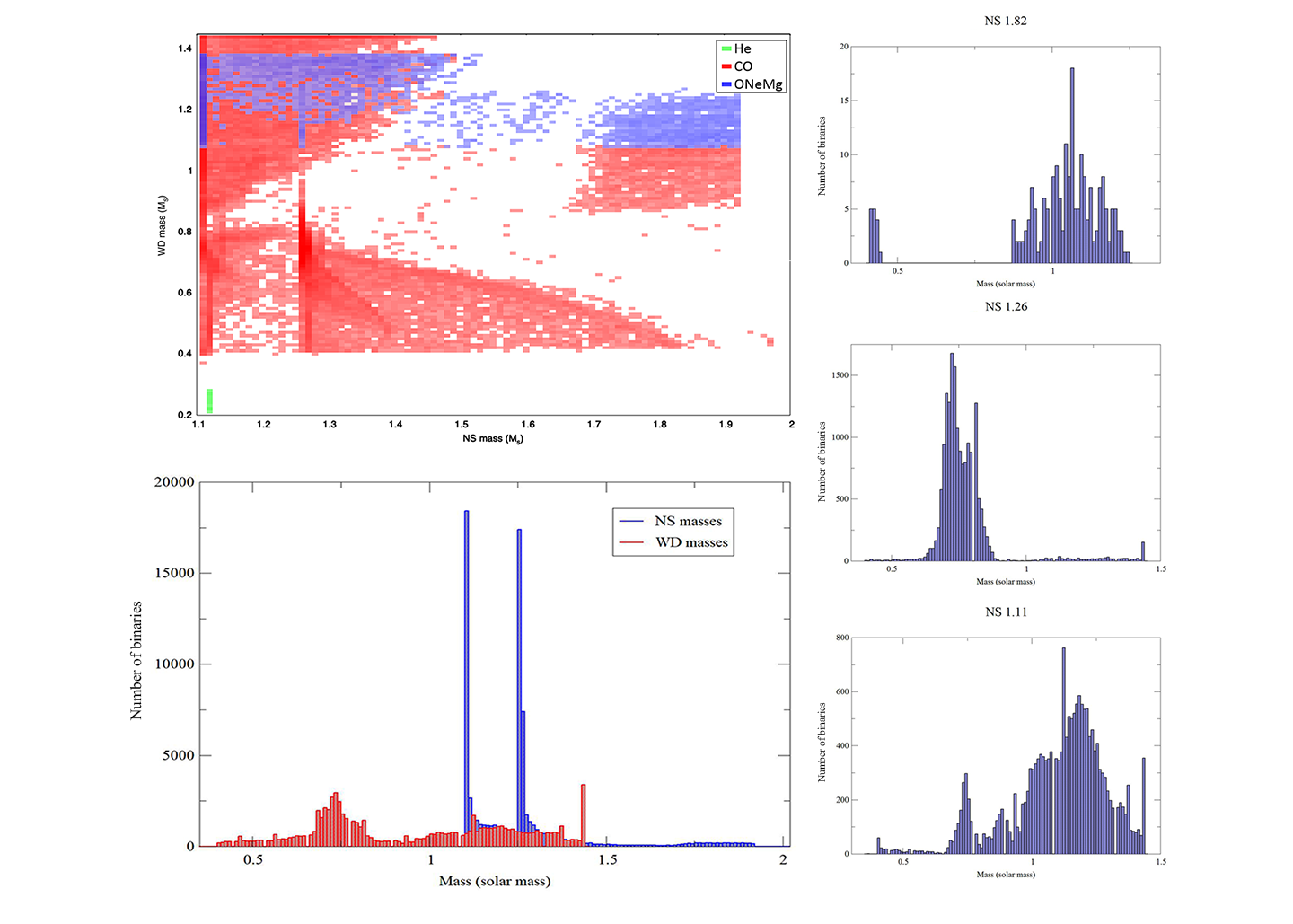}
\caption{The demographics ensued from running the StarTrack code and collecting the NS-WD binaries. The top left figure shows a heat map of NS-WD binary mass components with the different colors representing the dominant nuclear composition. The bottom left figure shows the absolute mass distribution for all WDs and NSs irrespective of their particular companion. On the right, we show the WD mass distribution for select neutron star masses where we find peaks in the distribution, $M_{\text{ns}} = 1.11 \text{, and } 1.26$  and an island of binaries centered around $1.82 M_\odot$.}
\label{fig:popsyn}
\end{figure}

We focus on the population demographics for NS-WD compact binaries. Figure~\ref{fig:popsyn} shows the mass distributions for the resulting NS-WD binaries. We see strong peaks in the mass distributions for the NS at $M_{\text{ns}} = 1.11 \text{ and } 1.26 M_\odot$. Additionally, we can see large zones where few binaries exist, and an island of binaries formed in the range of NS from $M_{\text{ns}} = 1.70 \text{ to } 1.90 M_\odot$ and WD from $M_{\text{ns}} = 0.90 \text{ to } 1.30 M_\odot$. The WD distribution is less strongly peaked than the NS distribution; however, peaks can still be seen in histograms in the figure.

\subsection{\label{subsec:UMT}Unstable Mass Transfer}
Once the double-compact binaries are formed, they enter a phase of their lives lasting millennia. The evolution of their orbits is largely described by the emission of gravitational waves, and NS-WD binaries are primarily emitting in the mHz frequency band. Future planned GW detector missions will open up this band for observational opportunities. Not only will these GW emissions aid in getting better values of binary populations but will also allow for dual observing the merger of NS-WD binaries in GWs and in more traditional electromagnetic spectra. Once the GW emission has driven the WD to the critical Roche-Lobe Overflow (RLOF) separation, mass from the WD begins to transfer to the NS companion. Mass transfer typically occurs via one of two different methods. If the mass transfer occurs on a time scale proportional to the GW timescale, it is called ``stable''. Stable mass transfer typically requires that as the WD loses mass the binary orbit grows such that the WD remains at the critical RLOF separation. A well-known fit of the RLOF radius as a function of the binary mass ratio, $q=m_{\text{ns}}/m_{\text{ns}}$, and orbital separation, $a$, is given by:
\begin{eqnarray}
    \label{eq:mt1}
    R_{\text{RL}} = a r_{\text{RL}}(q) = a \frac{0.49 q^{2/3}}{0.6 q^{2/3}+\ln{(1+q^{1/3})}}\text{,}
\end{eqnarray}
where $R_{\text{RL}}$ is the Roch-Lobe radius \citep{Eggleton1983}. When mass is transferred from the WD to the NS, the orbital separation increases. The stability of mass transfer is maintained as long as the Roche-Lobe remains filled, that is, the timescale for the increasing of the WD Roche-Lobe radius is equal to the change in the WD radius, $\dot{R}_{\text{wd}} = \dot{R}_{\text{RL}}$. NS-WD binaries that survive the onset of mass transfer and become stable, evolve into long-living ultra-compact X-ray binaries, UCXBs \citep{Bobrick2017,Savonije1986}, after a short phase as ultra-luminous X-ray sources, ULXs \citep{Bildsten2004}. 

If the WD radius grows much faster than the Roche-Lobe radius, $\dot{R}_{\text{wd}} > \dot{R}_{\text{RL}}$, then the mass transfer quickly becomes unstable and the WD will be tidally disrupted. This leads to a merger that occurs on the order of the orbital timescale of the pre-merger binary system. To find the critical-mass ratio for stability, one needs to determine the value of $q$ for which the following inequality is true:
\begin{eqnarray}
\label{eq:unstablecriteria}
     \frac{\dot{R}_{\text{wd}}}{R_{\text{wd}}} &>& \frac{\dot{R}_{\text{RL}}}{R_{\text{RL}}}\nonumber\\
     \frac{d\ln R_{\text{wd}}}{d \ln m_{\text{wd}}} \frac{\dot{m}_{\text{wd}}}{m_{\text{d}}}&>&\frac{\dot{R}_{\text{RL}}}{R_{\text{RL}}}\nonumber\\
     \frac{d\ln R_{\text{wd}}}{d \ln m_{\text{wd}}}\frac{\dot{m}_{\text{wd}}}{m_{\text{d}}}&>&\frac{\dot{a}}{a} +\frac{d\ln r_{\text{RL}}}{d \ln q}\frac{\dot{q}}{q}\text{.}
\end{eqnarray}

The change in the orbital separation with time is determined by conserving the change in total angular momentum of the orbit. We take into account mass loss from the orbit, GW emissions, and mass transfer between the two components with the following form:

\begin{eqnarray}
    \frac{\dot{a}}{a} &=& \frac{\dot{\mathcal{J}}_{\text{gw}}}{\mathcal{J}_{\text{orb}}}\frac{2 \mathcal{J}_{\text{orb}}}{\left(\mathcal{J}_{\text{orb}} +\mathcal{J}_{\text{d}}\right)} -\left[2 \mathcal{J}_{\text{orb}} -2(1-\beta)\left(q \mathcal{J}_{\text{orb}} + \frac{m_{\text{wd}}}{m_{\text{d}}}\mathcal{J}_{\text{d}}\right)+\frac{d\ln r_{\text{d}}}{d \ln q}\mathcal{J}_{\text{d}}-\beta \frac{q}{q+1}\mathcal{J}_{\text{orb}}\right]\frac{\dot{m}_{\text{wd}}}{m_{\text{wd}}\left(\mathcal{J}_{\text{orb}} +\mathcal{J}_{\text{d}}\right)}\text{,}
\end{eqnarray}
where $\mathcal{J}$ is the angular momentum for each component labeld with the subscript; the subscript $\text{d}$ denotes the disk; the subscript $\text{orb}$ denotes the orbit; $\beta$ is a parameterization for the fraction of material that is ejected from the local system; $r_{\text{d}}$ is the radial fraction of the disk assumed to be $R_{\text{d}} = r_{\text{d}}a$. 

By solving for when the left-hand side of Eq. \ref{eq:unstablecriteria} is equal to the right-hand side for different NS masses, we find the line of maximum stability. When the mass of a WD companion of a NS is larger than the maximum stability line, we find that even under unrealistically unfavorable conditions{\textemdash}where the accreting mass does not form a disk, rather directly accretes onto the NS{\textemdash}mass transfer becomes unstable for mass ratios greater than about $q \geq 0.4$. With our more favorable and realistic conditions, when the accreting mass does form a disk, we find that mass transfer becomes unstable for mass ratios greater than about $q > 0.2$. For more detailed analysis, refer to \cite{Fryer1999a} discussing the intricate stability line for the similar BH-WD systems.

Other studies of the stability of mass transfer from the WD to the NS additionally take into account the structure, composition, and temperature of the WD as well as the mode and amount of mass and angular momentum loss from the binary during the accretion phase. They have found the critical mass ratio to be $q\approx 0.40-0.53$ \citep{Verbunt1988,Paschalidis2009}. More recent arguments suggest that winds from the accreting stream are far more important to the stability, determining a much smaller $q_{\text{crit}}\gtrsim 0.20$ \citep{Bobrick2017}, which is in line with our results.
\iffalse 
\begin{figure}[h]
    \includegraphics[width=0.95\columnwidth]{Mns_Mwd_stable_q_withq0.5.jpg}
    \caption{The plot shows the lines of stability for binary systems during mass transfer from the WD to the NS. Any NS-WD binary with components which fall beneath a given line would continue with stable mass transfer for the existence of the binary. $\beta$ is the mass fraction that is ejected from the system. The black curves represent mass transfers into a disk around the NS. The red curve represents if mass transfers directly onto the NS. The grey dotted line is the $q=0.5$ line for reference.}
    \label{fig:unstable}
\end{figure}
\fi

Combining peak values in the population distributions with the more conservative value for the critical mass ratio, $q\geq 0.40$, we chose to simulate a variety of binary systems with NS masses 1.10. 1.25, 1.40, and 1.80 $M_\odot$. The WD mass is chosen for a given binary as long as the mass ratio should conservatively result in unstable mass transfer and tidal disruption of the WD. Table~\ref{table1} shows the selection of key binary systems and information regarding their state at the start of mass transfer.

\begin{table}[h]
    \begin{tabular}{|c|c|c|c|c||c|c|c|c|c|}
    \hline
    $M_{\text{wd}}$ & $M_{\text{ns}}$ & $q$ & $a_{\text{sep}}$ & $P$ & $M_{\text{wd}}$ & $M_{\text{ns}}$ & $q$ & $a_{\text{sep}}$ & $P$\\
    \hline
    0.50 & 1.10 & 0.45 & 3.11e9 & 74.7 & 1.00 & 1.40 & 0.71 & 1.58e9 & 22.2\\
    \hline
    0.50 & 1.25 & 0.40 & 3.21e9 & 75.1 & 1.00 & 1.80 & 0.56 & 1.68e9 & 22.5\\
    \hline
    0.75 & 1.10 & 0.68 & 2.15e9 & 40.2 & 1.25 & 1.10 & 1.14 & 8.95e8 & 9.53\\
    \hline
    0.75 & 1.25 & 0.60 & 2.22e9 & 40.5 & 1.25 & 1.25 & 1.00 & 9.21e8 & 9.65\\
    \hline
    0.75 & 1.40 & 0.53 & 2.29e9 & 40.7 & 1.25 & 1.40 & 0.89 & 9.46e8 & 9.74\\
    \hline
    1.00 & 1.10 & 0.91 & 1.49e9 & 21.8 & 1.25 & 1.80 & 0.69 & 1.00e9 & 9.93\\
    \hline
    1.00 & 1.25 & 0.80 & 1.54e9 & 22.0 & & & & &\\
    \hline
    \end{tabular}
    \caption{In this table, we provide the following for the binary configurations selected for study: component masses, $M_{\text{wd}}$ and $M_{\text{ns}}$; the mass ratio $q = M_{\text{wd}}/M_{\text{ns}}$; the orbital separation at RLOF, $a_{\text{sep}}$; and the orbital period at this separation, $P$. All masses are given in solar mass, $M_\odot$. The orbital separation is given in cm, and the orbital period is given in seconds. Due to the age of the binaries at merger, all orbits are assumed to have circularized, i.e., the eccentricities of the orbits are negligible.}
    \label{table1}
\end{table}
\subsection{Initial Disk Conditions}
We are interested in binary systems shown in Table~\ref{table1}. Once the WD undergoes unstable mass transfer at $a_{\text{RLOF}}$, the WD quickly forms a disk, as it is tidally disrupted over a few orbital periods, with characteristic dimensions proportional to the circularization radius $R_c = (1+q)^{-1}$. $R_c$ is defined as the semi-major axis of a point of mass $M_{WD}$ orbiting the central NS of mass $M_{NS}$, with an angular momentum equal to that of the initial binary at the beginning of mass transfer.

High-resolution hydrodynamic simulations of the binary systems are required to get a detailed description of the WD disruption and subsequent disk formation. Previous works have investigated these systems, or similar systems with a WD component, with limited resolutions \citep{Fryer1999a,Paschalidis2011}.
As a prelude to high-resolution simulations of a the suite of NS-WD binaries, we instead must adopt a flexible analytic description for the initial disk density and temperature.

Here, we adopt the initial density profile described by \cite{Margalit2016}. The disk-surface density, related to the mid-plane density by $\Sigma = 2 h r \rho$, is given by 
\begin{eqnarray}\label{eq:rhoprof}
     \Sigma_0 (r) = \mathcal{N}(m,n) \frac{M_{wd}}{2 \pi R_d^2}\left(\frac{r}{R_d}\right)^m\left[1 + \frac{m+2}{n-2}\frac{r}{R_d}\right]^{-(m+n)}\text{,}
\end{eqnarray}
where $R_d = \mathcal{R}(m,n) R_c$ is the characteristic disk radius at which the radial mass peaks; r is the radius centered on the central object; and $\mathcal{N}$ and $\mathcal{R}$ are constants $\lesssim 1$, calculated assuming that the mass and angular momentum are conserved during the tidal disruption. Here we have chosen $(m, n) = (2, 7)$. 

The initial temperatures for the disk were approached in a few different ways. Initially, we applied the temperature profile as described in \cite{Margalit2016}. At large radii in the disk, the gas pressure dominates over the radiation pressure, and, at closer radii, the radiation pressure is dominant. Thus, the disk temperature is fit by 
\begin{eqnarray}
\label{eq:t1}
  T(r) = 
    \begin{cases}
      (\mu m_p/k_B)h^2\Omega_k^2r^2 & \text{, gas}\\
      [(3/2a)h\Omega_k^2 r\Sigma]^{1/4} & \text{, radiation}
    \end{cases}
\end{eqnarray}
where $\mu$ is the mean molecular weight; $m_p$ is the mass of the proton; and $a$ is the Boltzman constant. 

For the second temperature prescription, we applied a radiation-dominated temperature profile that scales with the density of the disk, given by
\begin{eqnarray}
\label{eq:t2}
     T(r) &=& \left(\frac{\rho}{1.1\times 10^{-11}}\right)^{1/4} \left(\frac{3}{a_r}\right)^{1/3}\text{,}
\end{eqnarray}
where $a_r$ is the radiation density constant.

A third prescription was applied specifically to assume low-entropy material in the initial disk with no attention to heating during the tidal disruption. Taking into consideration the age of the binaries prior to merger, the WD material would have had $\sim 10^9$ years to cool. Thus, we expect the entropy of the material that will form the disk to be very low. As such, we designed a temperature prescription which would provide the desired low-entropy conditions. This third temperature prescription was initially designed to be isoentropic; however, a scaling factor was introduced to allow for radial variation,
\begin{eqnarray}
\label{eq:t3}
     T_{\text{MeV}}(r) &=& \left(\frac{A(r) s \rho}{5.2\times 10^{8}}\right)^{1/3} \text{,}
\end{eqnarray}
where $A(r)$ and $s$ are the radially dependent scaling factor and the desired entropic parameter, respectively. 

\begin{figure}[h]
    \centering
    \includegraphics[width=0.75\textwidth]{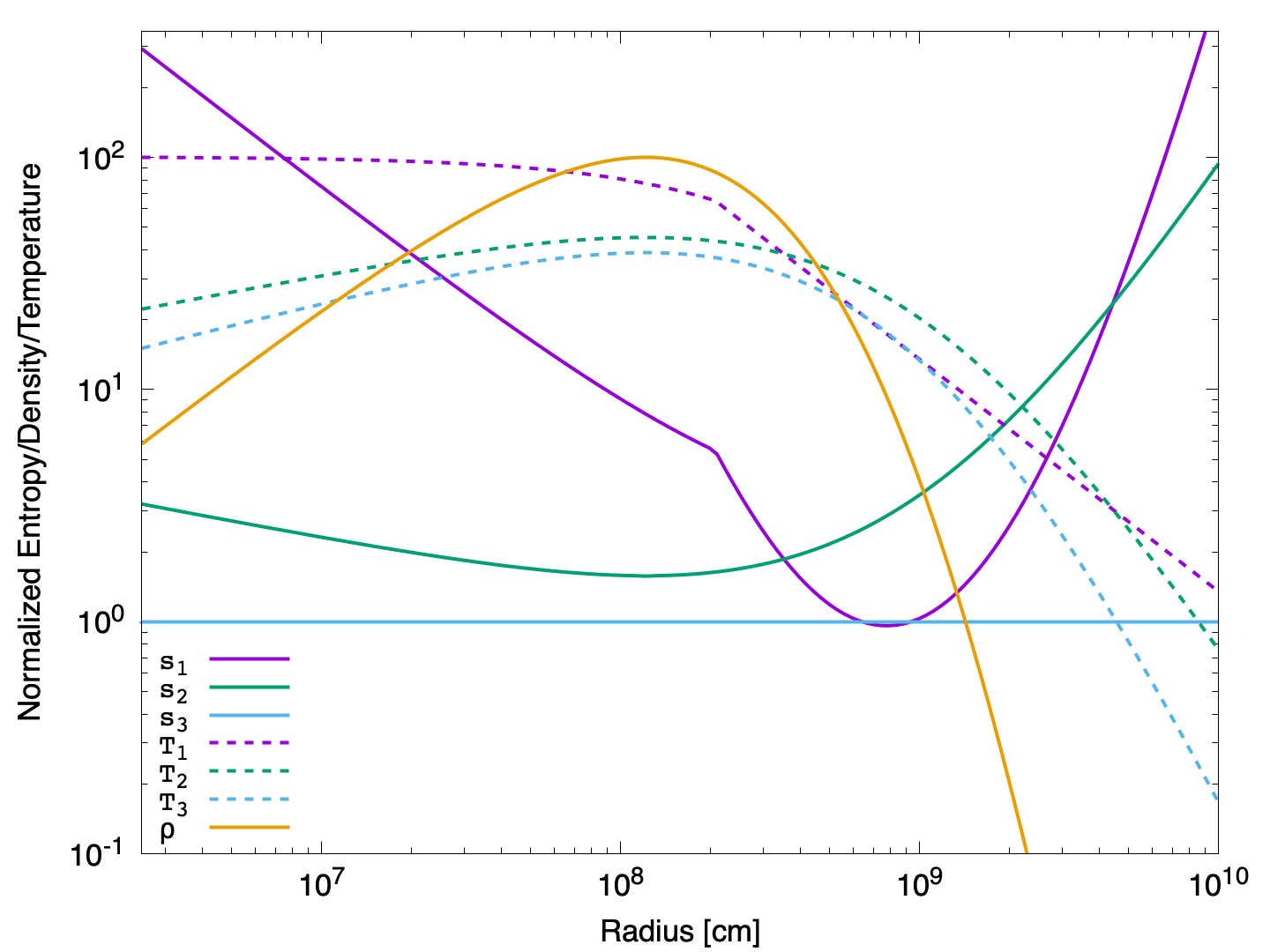}
    \caption{The different profiles for entropy for a $1.25 M_\odot$ NS and $1.00 M_\odot$ WD binary. The prescription is specified by the subscript, 1, 2, and 3. These refer to Eqs. \ref{eq:t1}, \ref{eq:t2}, \ref{eq:t3}, respectively. For prescription 1, the entropy of the material is high in inner and outer regions, dipping to its lowest point around $10^{9}$ cm. For prescription 2, the entropy decreases by an order of magnitude in the inner region; however, the middle and outer regions (where the bulk of the WD matter is) still have high entropies. In fact, there is a region where the entropies are higher than in  prescription 1. Prescription 3 can be tuned to have a desired entropy, and scaled as necessary, although here the entropy is assigned to be a constant. The temperatures are normalized to the maximum from prescription 1, $T_{\text{max}}=100$. The density profile is the same for each prescription, normalized to $\rho_{\text{max}}=100$.}
    \label{fig:entropies}
\end{figure}

The entropy of the initial disk material varies greatly with each of the above temperature prescriptions, as seen in Figure~\ref{fig:entropies}. In the following sections, we discuss the disk evolution for each of these conditions, in order to capture different potential effects from the tidal disruption. 

\section{Disk accretion and Wind Ejecta Model}\label{sec:accmodel}

The 2D structure of geometrically thin, non-self gravitating, axially symmetric accretion discs can be split into a $1+1$ structure corresponding to a hydrostatic vertical configuration and radial quasi-Keplerian viscous flow. The two 1D structures are coupled through the viscosity mechanism transporting angular momentum and providing the local release of gravitational energy.
\subsection{Disk Equations} 
We consider the accretion disks in this work to be geometrically thin, $z(r) < r$. This allows the equations of the accretion disk to be described by the vertically-integrated continuity equations. The thin disks are axisymmetric but not stationary. We use cylindrical coordinates $(r, \phi, z)$ and define the surface density of the disk, $\Sigma$, to be
\begin{eqnarray}
     \Sigma = \int_{-\infty}^{+\infty} \rho dz \approx 2 h \rho\text{,}
\end{eqnarray}
where $\rho$ and $h$ are the density and scale-height factor at the mid-plane of the disk. The conservation equation of mass is simplified because of the thin and axisymmetric approximations, i.e., $\frac{\partial}{\partial \phi} = 0$ and $\frac{\partial}{\partial z} = 0$; thus, we can write the equation as
\begin{eqnarray}\label{eq:masscont}
     \frac{\partial \Sigma}{\partial t} + \frac{1}{r} \frac{\partial}{\partial r} (r v_r \Sigma) + \Sigma_{\text{ext}} &= 0 \text{,}
\end{eqnarray}
where $v_r$ is the radial fluid velocity and $\Sigma_{ext}$ is the term which accounts for any mass lost or gained to the system. 

In this approximation, we use the Newtonian gravitational potential, $\Phi = -GM/r$ , where $M$ is the mass of the central object. The rotational velocity, angular velocity, and specific angular momentum are thus
\begin{eqnarray}
     v_k = \sqrt{\frac{GM}{r}}\text{, } \Omega_k = \sqrt{\frac{GM}{r^3}} \text{, and } j = \sqrt{GMr}\text{.}
\end{eqnarray}

Since we assume the disk is in vertical hydrostatic-equilibrium, the gravitational force is counteracted by the force produced by the pressure gradient, $\frac{dP}{dz} = \rho g_z$, where $g_z$ is the vertical component of the accreting body gravitational acceleration,
\begin{eqnarray}
     g_z =\frac{\partial}{\partial z}\left[\frac{GM}{\sqrt{r^2 +z^2}}\right] \approx \frac{GMz}{r^3}\text{.}
\end{eqnarray}

Taking the condition of geometrical thinness of the disk, $h < 1$, and writing $dP/dz \sim P/hr$, the scale factor can be rewritten as
\begin{eqnarray}
     h \approx \frac{c_s}{v_k}\text{,}
\end{eqnarray}
where $c_s \equiv \sqrt{P/\rho}$ is the mid-plane isothermal sound speed. 
The angular momentum conservation is dictated by the equation
\begin{eqnarray}
     \frac{\partial \Sigma j}{\partial t} +\frac{1}{r} \frac{\partial(r\Sigma j v_r)}{\partial r} - \frac{1}{r} \frac{\partial}{\partial r}\left(r^3 \Sigma \nu \frac{d\Omega_k}{dr}\right) + \Sigma_{\text{ext}} = 0 \text{.}
\end{eqnarray}

This conservation equation reflects the fact that angular momentum is transported through the disc by a viscous stress $r\Sigma \nu d\Omega/dr$. $\Sigma_{\text{ext}}$ represents the sink of the transported angular momentum, mass transferred to the central object and the momentum ejected from the system. Since we assumed the orbital angular velocity is Keplerian, we obtain the diffusion equation for the surface density,
\begin{eqnarray}
     \frac{\partial\Sigma}{\partial t} = \frac{3}{r} \frac{\partial}{\partial r} \left[r^{1/2} \frac{\partial}{\partial r} \left(\nu \Sigma r^{1/2}\right)\right]\text{.}
\end{eqnarray}

When we compare this relation to the mass conservation equation and solve for the radial velocity induced by the viscous torque, we get
\begin{eqnarray}\label{eq:vr}
     v_r \approx -3\frac{ \nu}{r}\frac{\partial \ln{r^2 \nu \Sigma \Omega_k}}{\partial \ln{r}} \text{,}
\end{eqnarray}
where $\nu$ is the kinematic viscosity. The $\alpha$-prescription is a simple description of the accretion disk physics, yet has proven better and physically more reliable due to this simplicity. The accretion-driving viscosity has a magnetic origin; however, one can use an effective hydrodynamical description of the accretion flow. The hydrodynamical stress tensor is
\begin{eqnarray}
    \tau_{r\phi} = \rho\nu \frac{\partial v_\phi}{\partial r} = \rho\frac{d\Omega}{d \ln{r}}\text{,}
\end{eqnarray}
where $v_\phi$ is the azimuthal velocity. \cite{Shakura1973} proposed the famous prescription
\begin{eqnarray}
    \tau_{r\phi} = \alpha P\text{,}
\end{eqnarray}
where $P$ is the total thermal pressure and $\alpha \leq 1$. This leads to
\begin{eqnarray}
    \nu = \alpha c_s^2 \frac{d\ln{r}}{d\Omega} =\alpha c_s \frac{h}{r}\text{,}
\end{eqnarray}
Since we assume $\Omega$ is Keplerian, we get
\begin{eqnarray}
    \nu \approx \alpha c_s^2/\Omega_k\text{.}
\end{eqnarray}
Using this form of the viscosity, Eq.~\ref{eq:vr} can be written as
\begin{eqnarray}
     v_r = -3\alpha h^2 v_k \left(2 + \frac{\partial \ln{\Sigma}}{\partial\ln{r}}+ 2\frac{\partial \ln{c_s}}{\partial\ln{r}}\right) \text{.}
\end{eqnarray}
We adopt $\alpha = 0.1$, consistent with those measured by numerical simulations \citep{Hoshino2015,Hung2019}. We can now use this radial velocity to determine the accretion rate of material in the disk via
\begin{eqnarray}
    \dot{M}(r) = -2\pi r \Sigma v_r \text{.}
\end{eqnarray}

Finally, we need to determine the energy conservation in the disk. The general form of the energy conservation (thermal) equation can be found by combining the first law of thermodynamics with the continuity equation. The specific internal energy $u$ evolves according to the first law of thermodynamics as \begin{eqnarray}
    \frac{\partial q}{\partial t} &=& \Sigma T \left(\frac{\partial}{\partial t} + v_r \frac{\partial}{\partial r}\right) s = \Sigma \left(\frac{\partial}{\partial t} + v_r \frac{\partial}{\partial r}\right) u - c_s^2 \left(\frac{\partial}{\partial t} + v_r \frac{\partial}{\partial r}\right) \Sigma\text{,}
\end{eqnarray}
where $s$ is the specific entropy and $\partial q/\partial t$ is the total disk heating rate per unit area. The heating rate can be disentangled into heating and cooling terms, $q^+$ and $q^-$, respectively. Using Eq.~\ref{eq:masscont}, we can rewrite the energy conservation in the disk as
\begin{eqnarray}\label{eq:encons}
    \frac{\partial u}{\partial t} = \frac{1}{\Sigma}\frac{\partial q}{\partial t} - v_r\frac{\partial u}{\partial r} - c_s^2 \left[\frac{1}{r} \frac{\partial (r v_r)}{\partial r} + \frac{\Sigma_{\text{ext}}}{\Sigma}\right]\text{.}
\end{eqnarray}
The disk heating and cooling incorporated in the model we have employed here are the viscous heating rate, the nuclear heating rate, the cooling from neutrino emission, and the wind cooling rate. We note that radiative cooling is neglected because we have assumed that the time-scale for photon diffusion from the disk mid-plane is much longer than the viscous time. 
The total torque $\mathcal{T}$ in the disk is found by multiplying the hydrodynamical stress tensor by the ring length $(2\pi R)$ and averaging over the disk height,
\begin{eqnarray}
    \mathcal{T} = 2 \pi r \Sigma \nu r\frac{d\Omega_k}{d\ln{r}} = 3\pi \Sigma \nu j\text{.}
\end{eqnarray}
The viscous heating is proportional to the hydrodynamical stress tensor. In particular, the viscous heating rate per unit volume is 
\begin{eqnarray}
    Q^+_{\text{visc}} = - \tau_{r\phi}\frac{d\Omega}{d\ln{r}} = \frac{3}{2}\alpha\Omega_k P\text{.}
\end{eqnarray}
The viscous heating rate per unit surface is therefore
\begin{eqnarray}
    q^+_{\text{visc}} = \Sigma \nu \Omega_k^2 \left(\frac{\partial \ln{\Omega}}{\partial \ln{r}}\right)^2 = \frac{9}{4}\alpha\Sigma c_s^2\Omega_k\text{.}
\end{eqnarray}

Nuclear reactions provide a source of heating and cooling $q_{\text{nuc}}$, which contributes to the net heating rate $q_{\text{tot}}$. This term is obtained by summing the energy production rates for all isotopes, extracted from the nuclear network.

The energy emitted from the production of neutrinos has thus far been neglected, so we introduce a neutrino cooling rate per unit surface: 
\begin{eqnarray}
    q^-_{\text{neut}} = -1.9\times10^{25}(8.617\times10^{-11} T)^9/\rho\text{.}
\end{eqnarray}
This becomes increasingly important at higher temperatures, since it scales $\sim T^9$. In fact, as expected the neutrino cooling term is significant in the inner most radii, quickly falling off with radius. At the innermost simulated radius, ($2.5\times10^6$\,km), the neutrino-cooling peaks at early times, comparable to the $t_{\text{max}}$ values for each initial WD mass compiled in Table \ref{tab:bigtabC}. In relation to the viscous heating, the neutrino cooling peaks at 10\%, 11.5\%, 14.5\%, and 17.5\% of the heating rate, for 0.50, 0.75, 1.00, and 1.25 $M_{\text{wd}}$, respectively. This corresponds to neutrino luminosities from $4.5\times 10^{34}$ to $1.5\times 10^{35}$ erg s$^{-1}$. With such a prominent contribution in the inner regions of the disk, a more accurate neutrino-cooling prescription should be investigated in the future.

\subsection{Accretion-Disk Wind Model} 
The matter ejected from the surface of the disk is important not only because the wind is an effective means of cooling down the accretion disk, but it also is the dominant source of optical emission from a NS-WD merger event. The outflows from the disk are included in the $\Sigma_{\text{ext}}$ sink terms above. We are neglecting any effect in the angular momentum conservation equation from the wind. Thus, the wind ejecta are only included in the mass and energy conservation equations. 
The wind prescription is taken from \cite{Margalit2016}. However, we have implemented it in a slightly different form. Thus, we describe the method below for reproducibility. The wind requires that a couple of parameters be defined: the cooling efficiency, $\eta_w$, and the Bernoulli parameter, $Be_d$. The wind-cooling efficiency is related to the wind velocity by $v_w = v_k\sqrt{2\eta_w}$. $\eta_w$ is chosen to be on the order of one. This assumes that the winds are ejected with velocities close to the material's escape speed.

The Bernoulli parameter of the disk mid-plane is given by
\begin{eqnarray}
     Be_d = \frac{1}{2} \Omega_k^2 r^2 +\frac{1}{2} v_r^2 + u + c_s^2 - v_k^2\text{.}
\end{eqnarray}
\cite{Narayan1995} originally showed that for one-dimensional models, the Bernoulli parameter is generally positive, suggesting that when the normalized Bernoulli parameter, $b = Be_d/v_k^2$, is positive, the gas is able to flow adiabatically outward on a radial trajectory, i.e., the gas would reach infinity with a net positive energy. If the value is $\leq 0$, then the gas cannot escape to infinity. 

We cool the disk with wind ejection when the material in the mid-plane has a normalized Bernoulli parameter which exceeds a critical value, in our case $b_{\text{c}} = 0$. This wind prescription assumes that there is some mechanism which preferentially heats the matter near the surface of the disk. This results in an effective cooling of the matter in the mid-plane by ejecting the hot surface material, in-turn lowering the Bernoulli parameter to less than the critical value.

Implementing the prescription in the disk equations gives the mass loss rate via wind ejection
\begin{eqnarray}
\label{eq:diskmassloss}
     \Sigma_{\text{ext}} = \dot{\Sigma}_w = \eta_w \Sigma \frac{\Omega_k}{h} \mathcal{H}(b - b_{\text{c}})\text{,}
\end{eqnarray}
where $\mathcal{H}$ is the Heaviside function. The material at the surface of the disk is only unbound when $b > b_{\text{c}}$. 
The wind efficiency parameter, $\eta_w$, effectively determines the efficiency at which the material ejected can cool the disk. The specific energy carried away in the wind is described by the wind cooling rate by
\begin{eqnarray}
\label{eq:diskheat}
     \dot{q}_w = -\dot{\Sigma}_w \frac{v_w^2}{2} = -\eta_w \dot{\Sigma}_w v_k^2 \text{.}
\end{eqnarray}
As long as $b_{\text{c}}$ is less than the wind efficiency parameter, the wind loss can cool the disk. Since the timescale for wind-mass loss, $t_w \sim hr/v_w \sim h/\Omega_k$, is smaller than the dynamical and viscous timescales in the disk, we assume that the wind ejection is effectively instantaneous. After carrying away the energy, the ejected material has no further effect on the disk.

\subsection{Numerical Method}
We numerically evolve the disks by converting equations \ref{eq:masscont} and \ref{eq:encons} to their finite difference forms using a logarithmically spaced grid, $\left[2.5\times 10^{6},1 \times 10^{10}\right]$ cm. We solved the equations by closing the disk-evolution equations with the Helmholtz EoS \citep{Timmes1999}. The Helmholtz EoS accurately accounts for an electron-positron gas with an arbitrary degree of degeneracy and relativistic motion, an ideal gas of ions, and a Planckian distribution of photons. The nuclear-reaction rates are evaluated using the publicly 21-isotope alpha network utilized in the MESA code \citep{cococubed}; additional post-processing was computed with the 495-isotope Torch network \citep{cococubed}. The variable timestep between each iteration is calculated by the Courant condition
\begin{eqnarray}
     \Delta t &=& C(t) \text{min}\left[ \frac{\Delta r^2}{\nu}\text{,} \frac{\Delta r}{c_s}\text{,}\frac{u}{\Delta u/\Delta t}\right] \text{,}
\end{eqnarray}
where $C$ is a scaling factor initially set to $0.1$. However, during the first few trial runs, it was determined that the scaling factor needed further refinement at early times to fully resolve the disk settling within the first second. To that end, the scaling factor was given a logarithmic time dependence $C(t)$ from $1\times 10^{-4}$ to $1\times 10^{-1}$ over the first second of evolution, after which the value remains constant.

\section{Disk-Evolution Results}\label{sec:numresults}
The disk models, implemented with the particular numerical prescription outlined above, were performed for three distinct classes over the same suite of NS-WD binary systems. Each class takes advantage of the same density profile, Eq.~\ref{eq:rhoprof}. Class 1, hitherto referred to as the high-entropy models, uses Eq.~\ref{eq:t1} for the temperature profile in the disk. Eq.~\ref{eq:t2} was also simulated, but the end results were nearly the same as Eq.~\ref{eq:t1}. The main difference occurred in the early times ($t<1s$) as the disk settled out, with lower wind ejecta in the small radii and higher wind ejecta in the largest radii. Because there is so little material in these regions at early times, the dynamics is dominated by the middle region where the majority of matter is located. Class 2, or low-entropy models, uses Eq.~\ref{eq:t3} assuming the scaling factor $A(r) = 1$, since during the early time when the disk is settling out, the inner regions are heated quite effectively when more material flows in. Class 3, the efficient-wind models, uses Eq.~\ref{eq:t3} with an additional modification to the disk heat-transport equation. Eq.~\ref{eq:diskheat} was modified to be twice as effective at radiating away heat without modifying the mass-loss rate equation \ref{eq:diskmassloss}. 

The most interesting disk results can be found summarized in Tables~\ref{tab:bigtabA}-\ref{tab:bigtabC}. Figure \ref{fig:classes} visualizes the results for the accretion rates and the disk winds for each class. 

\begin{table}[h]
    \begin{tabular}{|c | c | c | c | c | c | c | c | c|}
    \hline
         $M_{\text{wd}}$ & $M_{\text{ns}}$ & $q$ & $M^\text{tot}_{\text{acc}}$ & $\dot{M}^\text{peak}_{\text{acc}}$ & $t_{\text{max}}$ & $\Delta t_{1/2}$ & $M_{\text{wind}}$ & $M_{\text{wind}}/M_{\text{wd}}$\\
    \hline
         0.50 & 1.10 & 0.45 & 7.730E-03 & 6.231E-04 & 7.517 & 23.86 & 0.108 & 21.71\% \\
         0.50 & 1.25 & 0.40 & 7.297E-03 & 5.732E-04 & 7.724 & 24.27 & 0.101 & 20.30\% \\
         0.75 & 1.10 & 0.68 & 1.673E-02 & 2.244E-03 & 4.254 & 14.34 & 0.225 & 30.11\% \\
         0.75 & 1.25 & 0.60 & 1.581E-02 & 2.064E-03 & 4.401 & 14.76 & 0.214 & 28.55\% \\
         0.75 & 1.40 & 0.53 & 1.504E-02 & 1.919E-03 & 4.514 & 15.08 & 0.203 & 27.16\% \\
         1.00 & 1.10 & 0.91 & 3.006E-02 & 6.406E-03 & 2.594 &  8.70 & 0.366 & 36.69\% \\
         1.00 & 1.25 & 0.80 & 2.858E-02 & 5.909E-03 & 2.690 &  9.03 & 0.353 & 35.39\% \\
         1.00 & 1.40 & 0.71 & 2.729E-02 & 5.515E-03 & 2.742 &  9.30 & 0.341 & 34.17\% \\
         1.00 & 1.80 & 0.56 & 2.466E-02 & 4.745E-03 & 2.912 &  9.86 & 0.313 & 31.36\% \\
         1.25 & 1.10 & 1.14 & 5.070E-02 & 1.903E-02 & 1.331 &  4.63 & 0.522 & 41.81\% \\
         1.25 & 1.25 & 1.00 & 4.855E-02 & 1.768E-02 & 1.409 &  4.79 & 0.512 & 41.01\% \\
         1.25 & 1.40 & 0.89 & 4.677E-02 & 1.659E-02 & 1.443 &  4.95 & 0.502 & 40.17\% \\
         1.25 & 1.80 & 0.69 & 4.305E-02 & 1.456E-02 & 1.563 &  5.28 & 0.476 & 38.11\% \\
         \hline
    \end{tabular}
    \caption{\textbf{Class 1: high initial entropy.} The high initial entropy produces a high wind-ejection rate from the very beginning of the simulation, blowing off a major fraction of the initial mass of the WD before a significant fraction of the material has a chance to accrete through the inner boundary. All mass units are given in solar masses, $M_\odot$, and all time units are in seconds, $s$. For each configuration, the table shows the following: the total mass accreted through the inner boundary at roughly one minute into the simulation, the peak accretion rate through the inner boundary ($25$ km), the time at which the peak accretion rate occurs, the total time that the accretion rate is larger than half the maximum rate, the total mass ejected through wind, and the percentage of the initial WD mass carried away by the wind.}
    \label{tab:bigtabA}
\end{table}
\begin{table}[h]
    \begin{tabular}{|c | c | c | c | c | c | c | c | c|}
    \hline
         $M_{\text{wd}}$ & $M_{\text{ns}}$ & $q$ & $M^\text{tot}_{\text{acc}}$ & $\dot{M}^\text{peak}_{\text{acc}}$ & $t_{\text{max}}$ & $\Delta t_{1/2}$ & $M_{\text{wind}}$ & $M_{\text{wind}}/M_{\text{wd}}$\\
    \hline
         0.50 & 1.10 & 0.45 & 8.058E-02 & 4.413E-03 & 3.670 & 12.67 & 0.084 & 16.89\% \\
         0.50 & 1.25 & 0.40 & 8.111E-02 & 4.261E-03 & 4.539 & 13.33 & 0.078 & 15.69\% \\
         0.75 & 1.10 & 0.68 & 1.204E-01 & 1.200E-02 & 1.887 &  7.01 & 0.143 & 19.14\% \\
         0.75 & 1.25 & 0.60 & 1.239E-01 & 1.227E-02 & 2.030 &  6.98 & 0.132 & 17.60\% \\
         0.75 & 1.40 & 0.53 & 1.264E-01 & 1.229E-02 & 2.228 &  6.97 & 0.123 & 16.44\% \\
         1.00 & 1.10 & 0.91 & 1.530E-01 & 2.858E-02 & 1.278 &  3.62 & 0.215 & 21.55\% \\
         1.00 & 1.25 & 0.80 & 1.596E-01 & 2.804E-02 & 1.263 &  3.94 & 0.195 & 19.51\% \\
         1.00 & 1.40 & 0.71 & 1.646E-01 & 2.822E-02 & 1.193 &  4.07 & 0.179 & 17.97\% \\
         1.00 & 1.80 & 0.56 & 1.733E-01 & 2.827E-02 & 1.523 &  4.19 & 0.152 & 15.28\% \\
         1.25 & 1.10 & 1.14 & 1.836E-01 & 6.985E-02 & 0.665 &  1.72 & 0.301 & 24.10\% \\
         1.25 & 1.25 & 1.00 & 1.923E-01 & 6.989E-02 & 0.689 &  1.84 & 0.268 & 21.47\% \\
         1.25 & 1.40 & 0.89 & 1.996E-01 & 6.915E-02 & 0.710 &  1.95 & 0.243 & 19.46\% \\
         1.25 & 1.80 & 0.69 & 2.132E-01 & 6.932E-02 & 0.738 &  2.12 & 0.198 & 15.89\% \\
    \hline
    \end{tabular}
    \caption{\textbf{Class 2: low initial entropy.} The low entropy in the disk reduces the amount of material ejected in the wind, particularly for the larger mass initial WDs. With more material remaining in the disk, the total accreted material is roughly an order of magnitude larger, the peak rate is roughly double for each configuration, and the peak occurs at a much earlier time. The table is structured the same as Table~\ref{tab:bigtabA}.}
    \label{tab:bigtabB}
\end{table}
\begin{table}[h]
    \begin{tabular}{|c | c | c | c | c | c | c | c | c|}
    \hline
         $M_{\text{wd}}$ & $M_{\text{ns}}$ & $q$ & $M^\text{tot}_{\text{acc}}$ & $\dot{M}^\text{peak}_{\text{acc}}$ & $t_{\text{max}}$ & $\Delta t_{1/2}$ & $M_{\text{wind}}$ & $M_{\text{wind}}/M_{\text{wd}}$\\
    \hline
         0.50 & 1.10 & 0.45 & 1.255E-01 & 6.527E-03 & 3.865 & 13.83 & 0.036 & 7.32\% \\
         0.50 & 1.25 & 0.40 & 1.252E-01 & 6.316E-03 & 4.813 & 14.36 & 0.034 & 6.93\% \\
         0.75 & 1.10 & 0.68 & 1.848E-01 & 1.767E-02 & 2.024 &  7.38 & 0.059 & 7.90\% \\
         0.75 & 1.25 & 0.60 & 1.888E-01 & 1.776E-02 & 2.030 &  7.48 & 0.055 & 7.44\% \\
         0.75 & 1.40 & 0.53 & 1.914E-01 & 1.773E-02 & 2.369 &  7.51 & 0.053 & 7.10\% \\
         1.00 & 1.10 & 0.91 & 2.303E-01 & 4.231E-02 & 1.289 &  3.69 & 0.085 & 8.50\% \\
         1.00 & 1.25 & 0.80 & 2.383E-01 & 4.102E-02 & 1.311 &  4.06 & 0.078 & 7.84\% \\
         1.00 & 1.40 & 0.71 & 2.441E-01 & 4.056E-02 & 1.268 &  4.27 & 0.073 & 7.37\% \\
         1.00 & 1.80 & 0.56 & 2.537E-01 & 3.974E-02 & 1.586 &  4.44 & 0.065 & 6.59\% \\
         1.25 & 1.10 & 1.14 & 2.684E-01 & 9.993E-02 & 0.654 &  1.76 & 0.114 & 9.19\% \\
         1.25 & 1.25 & 1.00 & 2.785E-01 & 1.000E-01 & 0.704 &  1.87 & 0.103 & 8.25\% \\
         1.25 & 1.40 & 0.89 & 2.886E-01 & 9.851E-02 & 0.752 &  1.94 & 0.094 & 7.57\% \\
         1.25 & 1.80 & 0.69 & 3.012E-01 & 9.678E-02 & 0.772 &  2.19 & 0.080 & 6.46\% \\
         \hline
    \end{tabular}
    \caption{\textbf{Class 3: low initial entropy and efficient wind.} The disk material initially had the same entropy profile as seen for the low-entropy models. However, the wind was twice as efficient at carrying away heat from the disk. This resulted in a much-reduced mass ejection throughout the entirety of the simulation. The table is structured the same as Table~\ref{tab:bigtabA}.}
    \label{tab:bigtabC}
\end{table}

\subsection{High Entropy}
When the material is initialized with the temperature profile determined by Eq.~\ref{eq:t1}, we find strong wind ejection for each binary configuration. The results from this class of models is summarized in Table~\ref{tab:bigtabA}. Due to the strong ejections, ranging from $21\text{-}42\%$ of the initial disk mass, the accretion rate through the inner boundary is suppressed. Though the peak accretion rates are low, the duration that the accretion rates are maintained is much longer between 5-24 seconds depending on the WD initial mass. This is in contrast to the other classes, which have much lower sustained accretion times. When comparing these accretion rates to the steady-state disk-accretion approximation $\dot{M} \approx \frac{M}{P/\alpha}$ where $P$ is the orbital period at merger, $\alpha$ is the viscous alpha parameter, and $M$ is the mass of the initial WD, the peak of these disks' accretion through the inner boundary reach close to these approximations. The times of peak accretion are close to the time of peak wind ejection. However, when analyzing the mass ejection, we find a significant fraction of all the wind material is blown off the disk at early times and at large radii, $t < t_{\text{max}}$ and $r > 10^8$ cm (see Figure~\ref{fig:classes}).

We ran a suite of simulations using the temperature profile from Eq.~\ref{eq:t2}. The results were in line with the results from the high-entropy models. The increased entropy region in the mid to large radii (see Figure~\ref{fig:entropies}) resulted in slightly larger wind ejecta and a very slight increase in the peak wind-ejecta time, but no significant changes in accretion rates. As such, the results are not summarized in their own subtable. 

\subsection{Low Entropy}
When the material is initialized with the temperature profile determined by Eq.~\ref{eq:t3}, we find the material ejected by the wind to be significantly reduced, particularly for the larger mass WDs. For example, for the NS-WD $1.10 M_\odot$-$1.25 M_\odot$ configuration, the low-entropy model ejected 24\% of the initial disk mass, compared with 41.8\% of the material for high-entropy models. The peak wind-ejection times were coincident with those for the high-entropy models; however, the dynamics and radial emission of the wind have changed substantially (see Figure~\ref{fig:classes}). The radial wind-ejection profiles shifted towards the inner radii, $r < 3\times 10^8$ cm. Most of the disk mass initially resides around $10^8$ cm. In the low-entropy and efficient models, this bulk mass experiences significantly less wind ejection from the beginning of the simulation.

The accretion rates for low-entropy are significantly higher than those for high entropy. The lower mass $0.50 M_\odot$ WD configurations saw a $\sim 700\%$ increase in their peak accretion rates, while the higher mass $1.25 M_\odot$ WD configurations increased by $\sim 350\%$. These higher accretion rates were sustained for half to a third of the $\Delta t_{1/2}$ seen for high entropy. These higher rates are a result of the lower wind rates, i.e., larger disk masses produce higher accretion rates. 

\subsection{Efficient Wind}
We consider that our wind-cooling mechanism might be underestimating the cooling efficiency for the amount of material being ejected. In order to investigate the effect of this assumption, we double the effective cooling rate without modifying the amount of ejected material. In doing so, we found the most significant changes to our ejected material results. The material ejected in the outer regions of the disk are significantly reduced for all the models, especially so for the higher WD mass models. This allows for more material to reach the inner regions of the disk, which in turn increases the wind in the inner regions as compared to the previous two classes. Since more material remains in the disk, we find the peak accretion rates increase for all models, while the timing of the accretion rate, $t_{\text{max}}$ and $\Delta t_{1/2}$ are similar to those of the low-entropy models. 

\subsection{Accretion Rates}
\begin{figure}[h]
\centering
  \includegraphics[width=.45\linewidth]{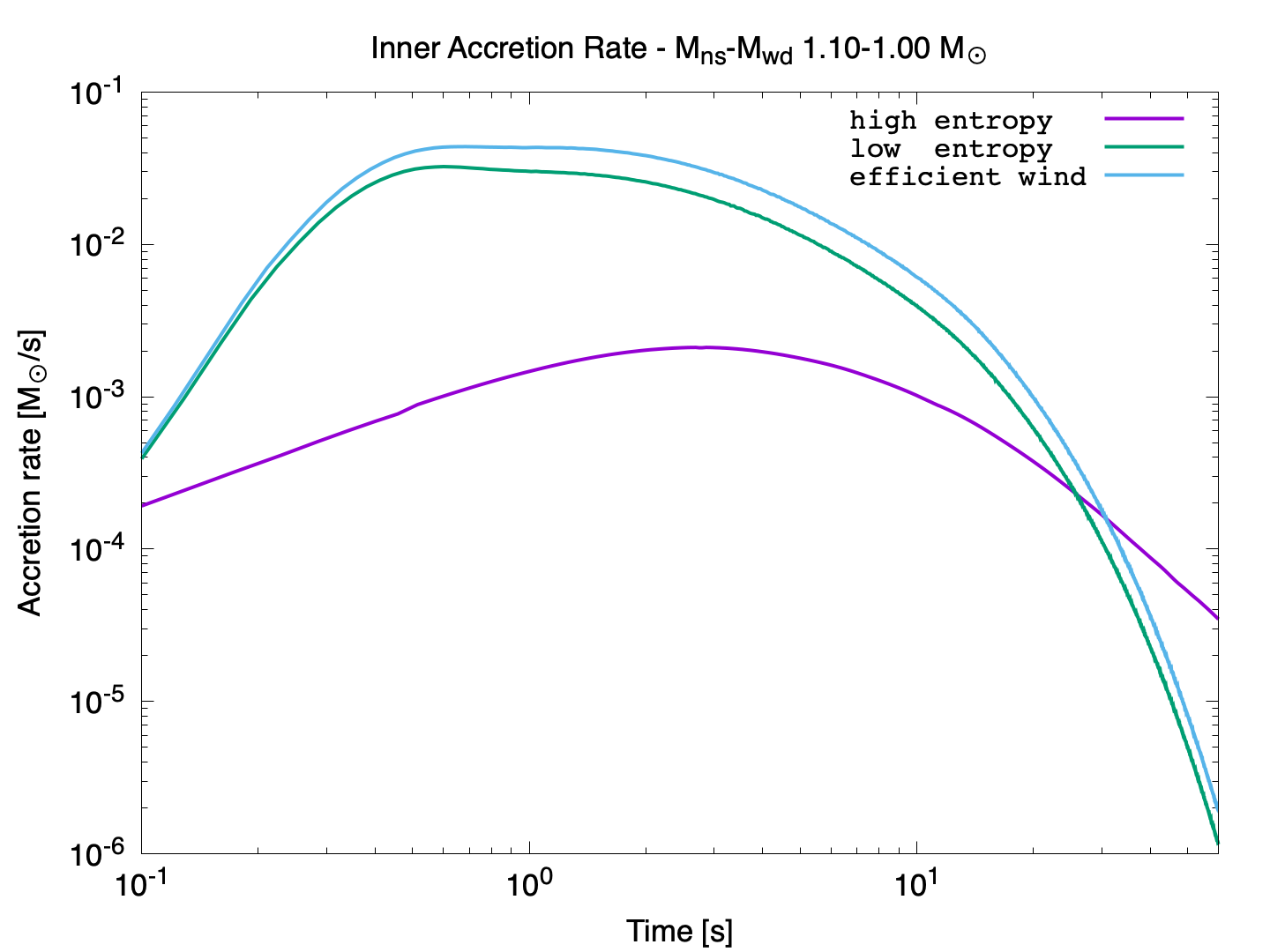}
  \includegraphics[width=.45\linewidth]{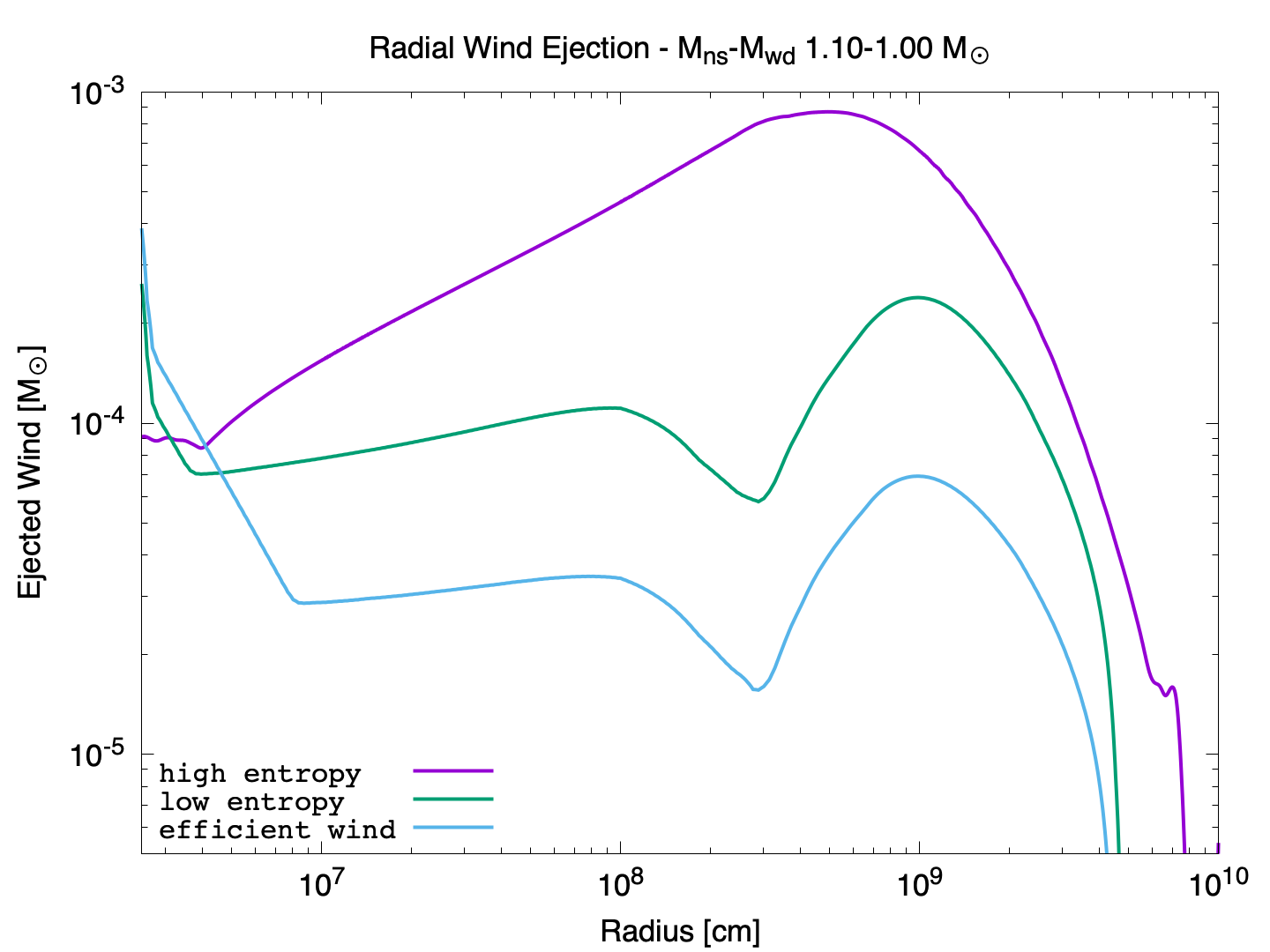}
\caption{The differences among the three classes of simulations for the accretion rate through the inner boundary (left) and for the wind-ejection radial profile (right). Here we specifically show the $M_{\text{ns}}$-$M_{\text{wd}}$ $1.25$-$1.00$ binary for each class.}
\label{fig:classes}
\end{figure}

The accretion rates are affected more strongly by the initial conditions of the disk than by the changes in the disk-wind efficiency. This is largely due to more material reaching the inner radii of the disk. In Figure~\ref{fig:classes}, we can see the differences for one select configuration for each class described above. There is a significant difference in the accretion rate for high-entropy models, as the disk winds begin carrying away material at a high rate from the very beginning of the simulations. The material in the disk is not able to reach the inner radii at sufficient densities for there to be high rates of accretion. When the wind efficiency is increased, we keep more of the material in the disk resulting in a higher rate of accretion, although still the same order of magnitude, but the accretion rate evolves nearly identically over time. It is important to note that this is the rate at which material flows through the inner boundary of our simulations (25 km), not to the surface of the NS. It is likely that the wind effects would be at least as strong in the inner-most region, and, thus, this accretion rate serves as an upper bound for the accretion onto the surface of the NS.

\subsection{Wind-Ejection Rates}
The wind ejected from the disk changes dramatically based on how the disk is initialized, suggesting that the proper implementation of the initial conditions is of paramount importance. Figure~\ref{fig:classes} shows the cumulative radial wind profile for each of the three classes. 
There is a significant difference in the wind ejection for the high-entropy models. The disk winds begin carrying away material at a high rate from the very beginning of the simulation, resulting in a smoother wind profile over the majority of the disk's length. The material is strongly ejected in the middle region of the disk, resulting in less material ever reaching the most central radii. When the low-entropy profile is applied, we can see that the amount of wind ejecta is strongly reduced in the middle region. In fact, we can see a local minimum, which corresponds to the point where the radial velocity of the disk changes from negative (inward) to positive (outward). When the wind efficiency is increased, we keep more of the material in the body of the disk, resulting in less material ejected until reaching the inner-most radii.

At radii larger than $r > 3 \times 10^8$ cm, we find that a large amount of material is ejected from the disk. However, this may be in large part due to the assumption that the material is not able to cool effectively through radiation emission. This may not be a good assumption at larger radii, where the densities are significantly lower and radiative emissions are more relevant.

\subsection{Nuclear Abundances}

\subsubsection{Disk Mid-line Abundances}
\begin{figure}[h]
\centering
  \includegraphics[width=.45\linewidth]{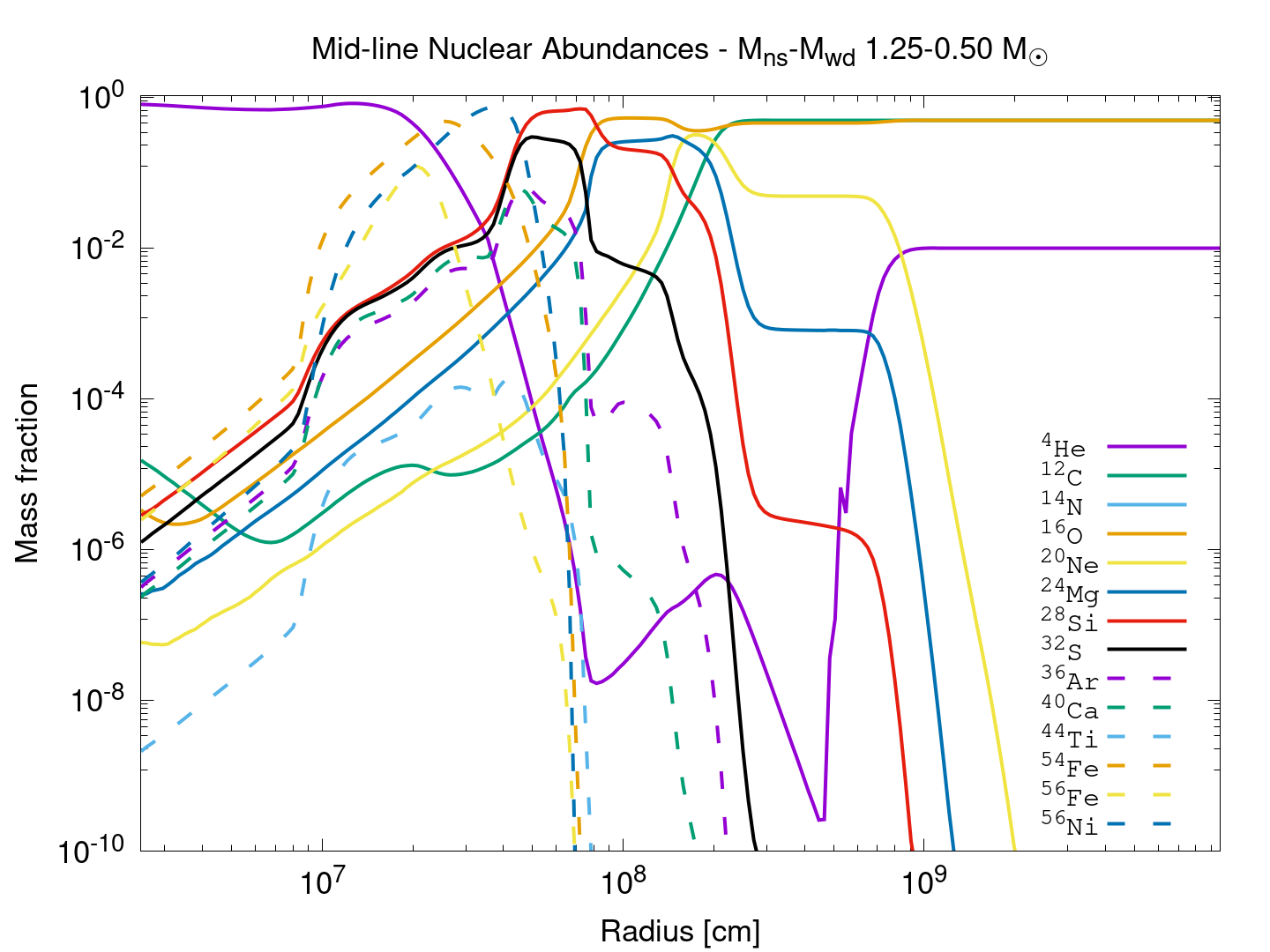}
  \includegraphics[width=.45\linewidth]{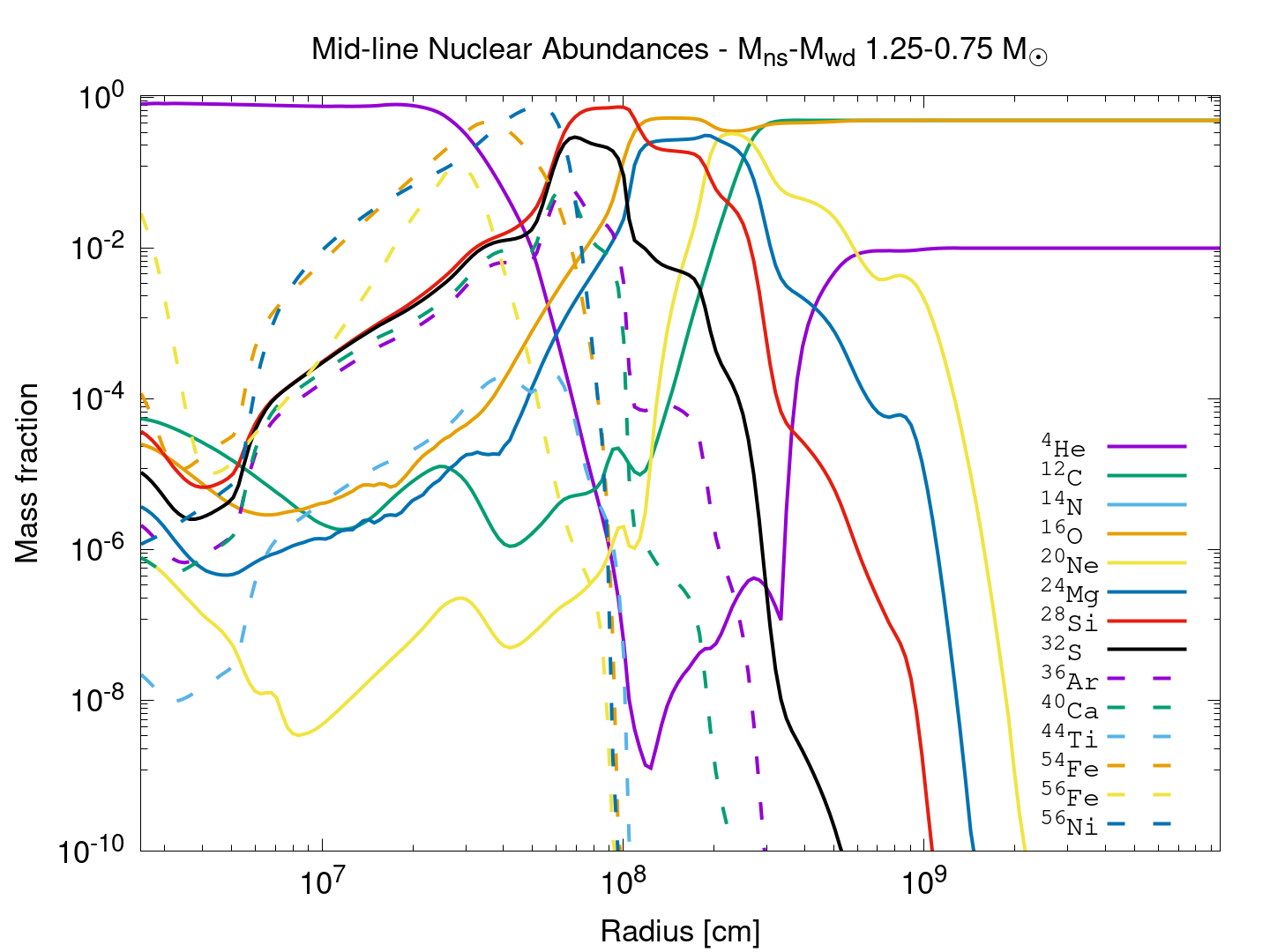}\\
  \includegraphics[width=.45\linewidth]{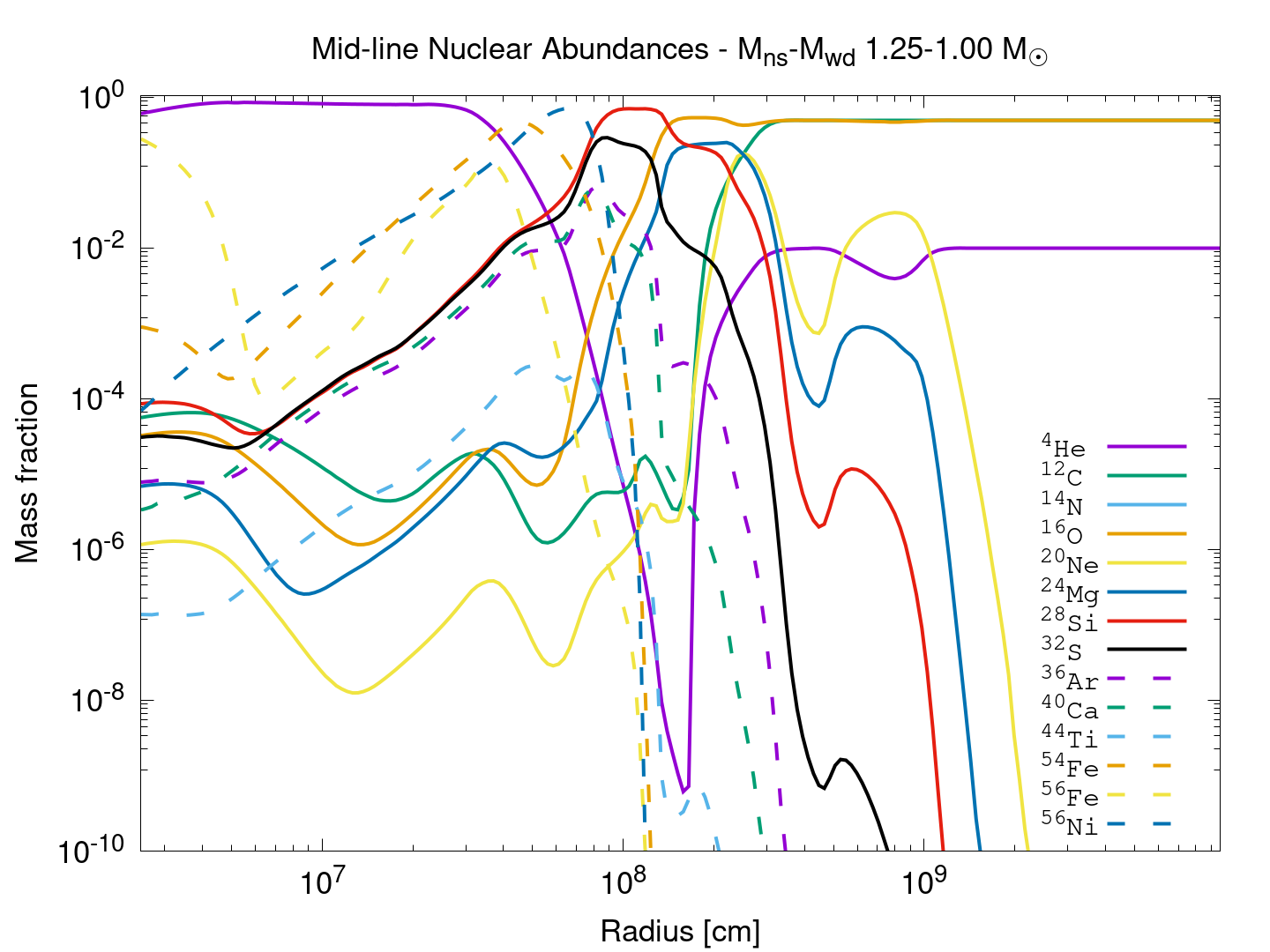}
  \includegraphics[width=.45\linewidth]{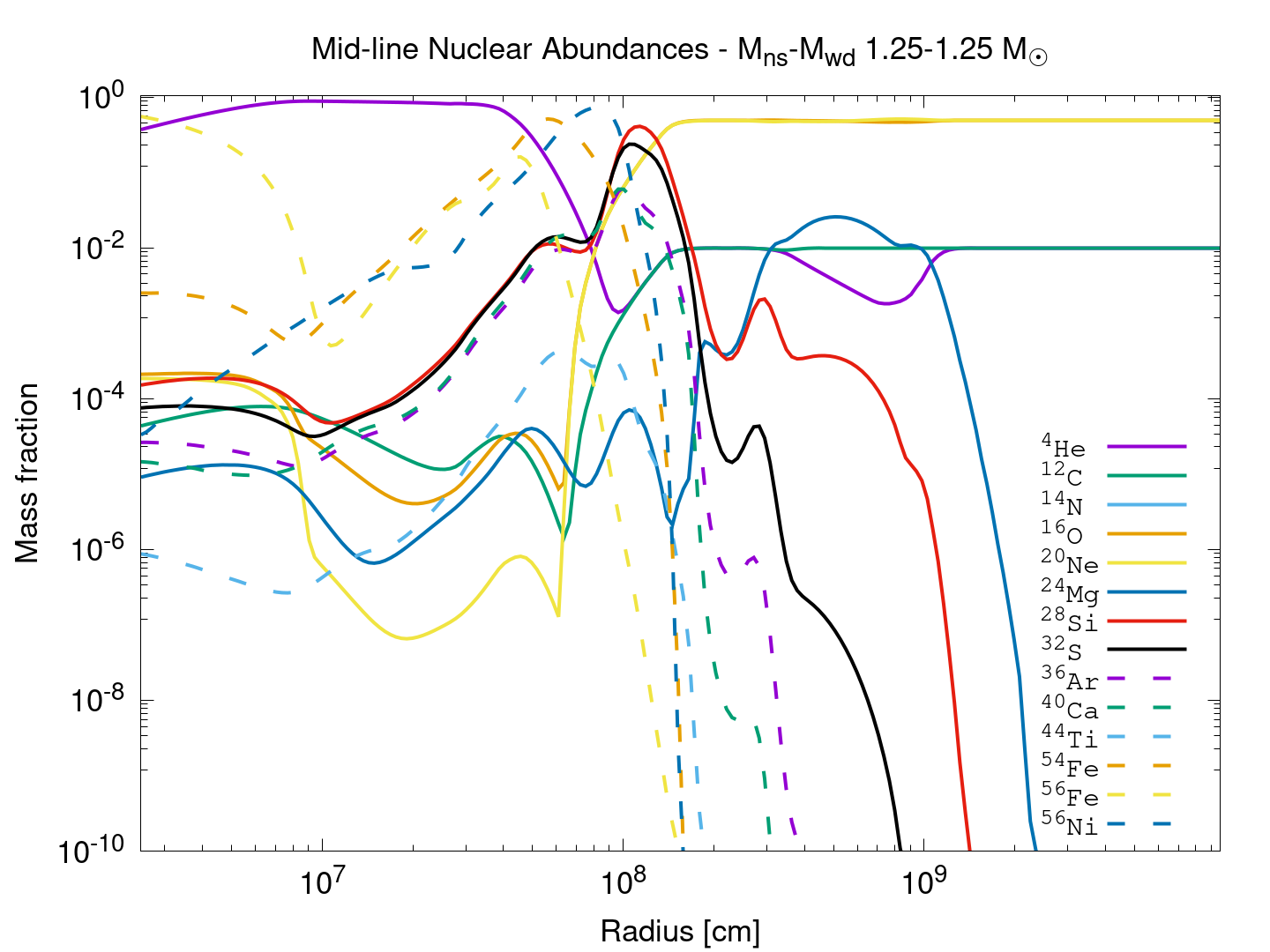}
  \caption{(Top Left) Initial WD mass $0.50 M_\odot$. (Top Right) Initial WD mass $0.75 M_\odot$. (Bot Left) Initial WD mass $1.00 M_\odot$. (Bot Right) Initial WD mass $1.25 M_\odot$. All companion NSs are $1.25 M_\odot$. Here we have the mid-line nuclear abundances at roughly $t=5$s for each disk model.}
  \label{fig:mlb}
\end{figure}

%Figure~\ref{fig:nuc} shows snapshots of the radial profile of the mass fractions for key isotopes in the midline of the disk for the different classes. 
The compositions of the disks end up taking a structure similar to massive stars. Successively heavier elements burn at decreasingly smaller radii, producing burning rings similar to the burning shells in stars. At a given radius, dependent on the mass of the initial WD and the NS (to a lesser degree), the temperature in the disk becomes sufficiently high ($T\sim 5\times 10^8$ K) to initiate the burning of the carbon-oxygen (doped with 1\% helium). For the 1.25 $M_\odot$ initial WD binaries, the initial composition is primarily oxygen-neon (doped with 1\% helium and carbon). Thus, the burning front for the largest disks requires the temperature to reach the neon ignition temperature ($T\sim 1.2\times 10^9$ K). As the radius decreases, further burning of the isotopes occurs, forming silicon, sulfur, argon, calcium, titanium, chromium and iron, and nickel. In the innermost radii, photodisintegration occurs, breaking the heavy elements into helium, protons, and neutrons. Figure~\ref{fig:mlb} shows the mid-line burning of material in the disk for four binary configurations from the efficient-wind models. We can see the effect of subsequently more massive disks on the location of the burning fronts.

All the disks form similar burning fronts with onset radii dependent on the different initial masses. The larger the initial mass of the WD, the farther out the burning fronts occur. The differences among Classes 1, 2, and 3 are more pronounced in the composition of the wind ejecta. 

\subsubsection{Disk-Wind Abundances}
\begin{figure}[h]
\centering
  \includegraphics[width=.45\linewidth]{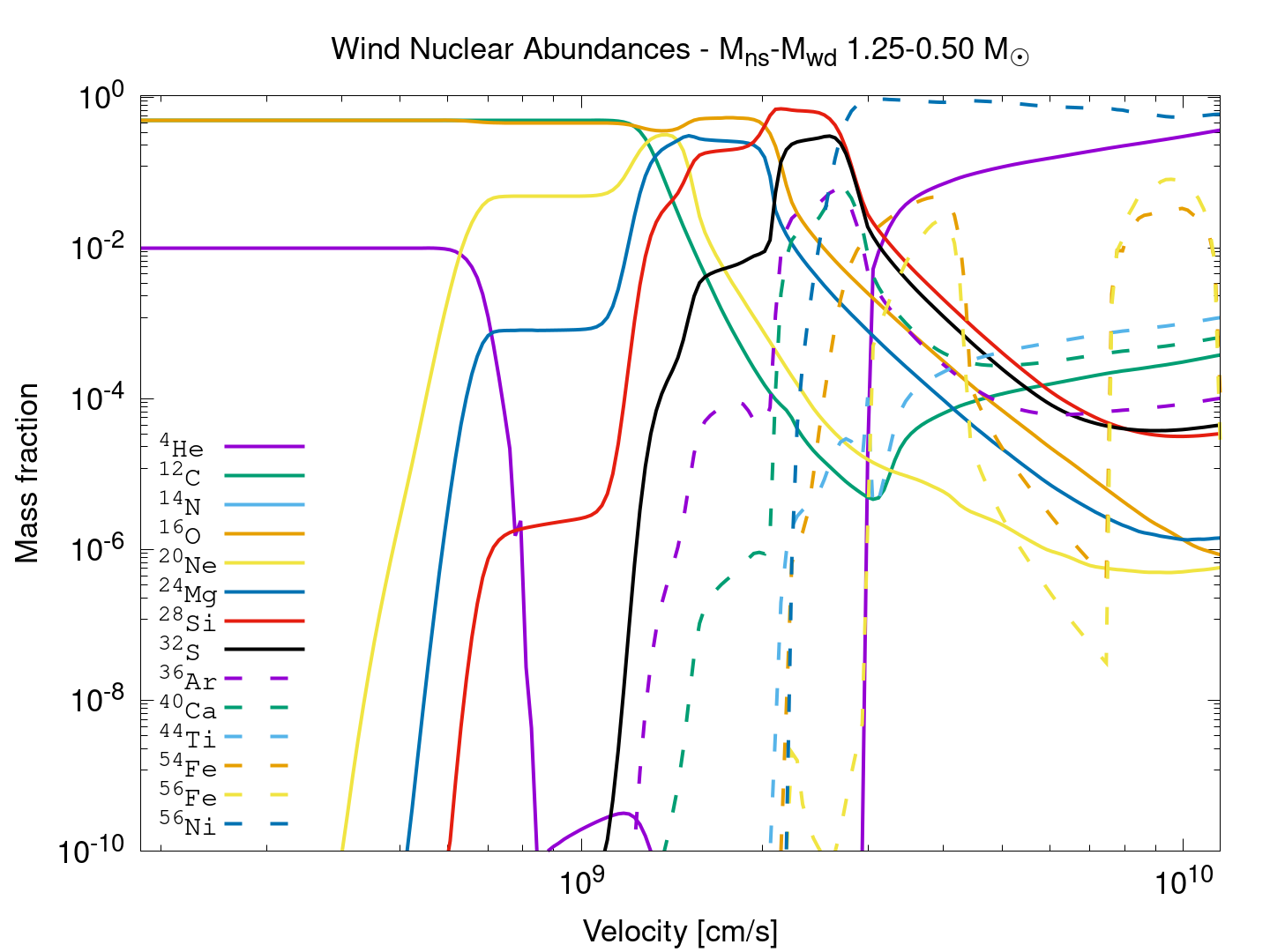}
  \includegraphics[width=.45\linewidth]{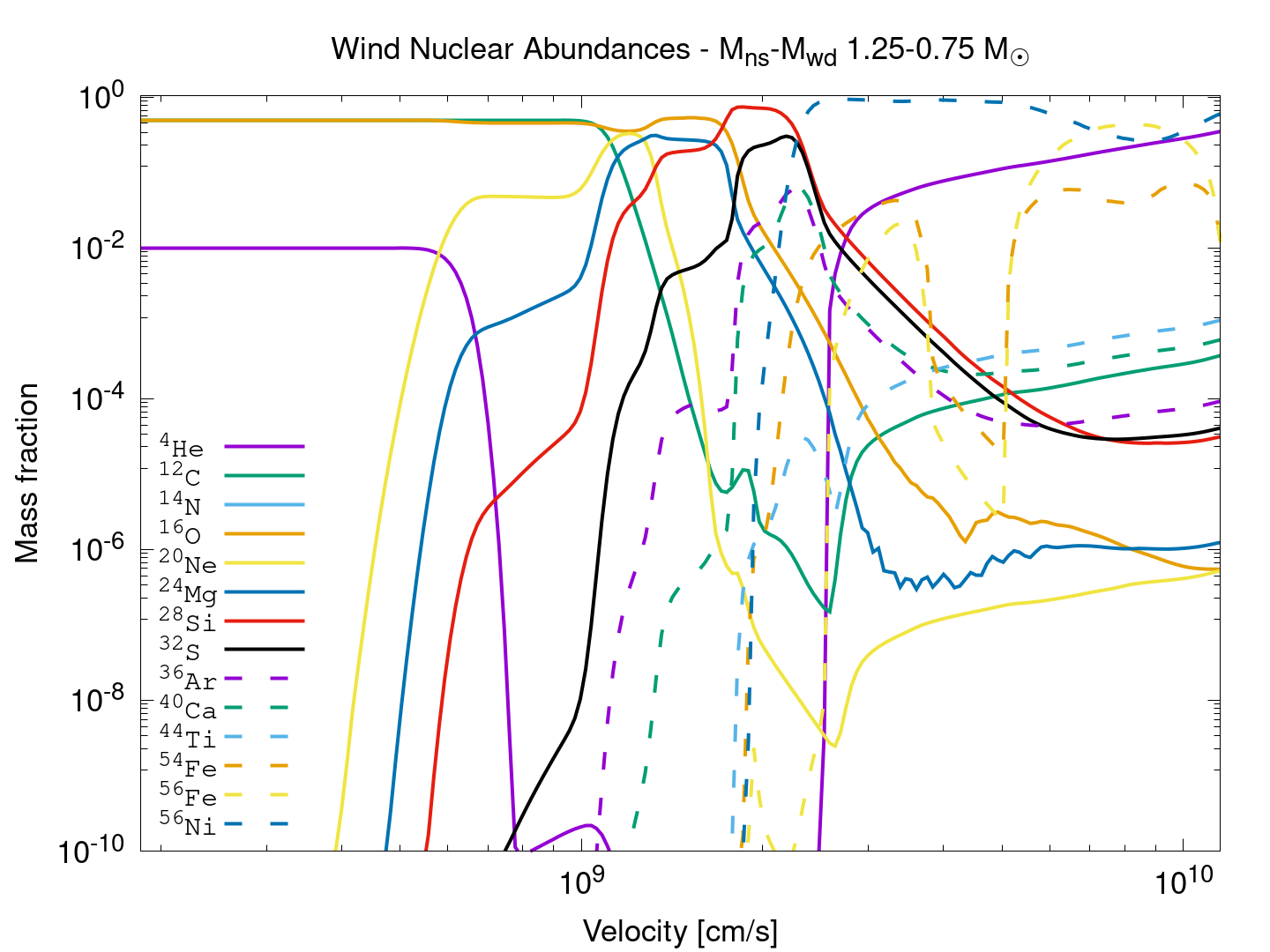}\\
  \includegraphics[width=.45\linewidth]{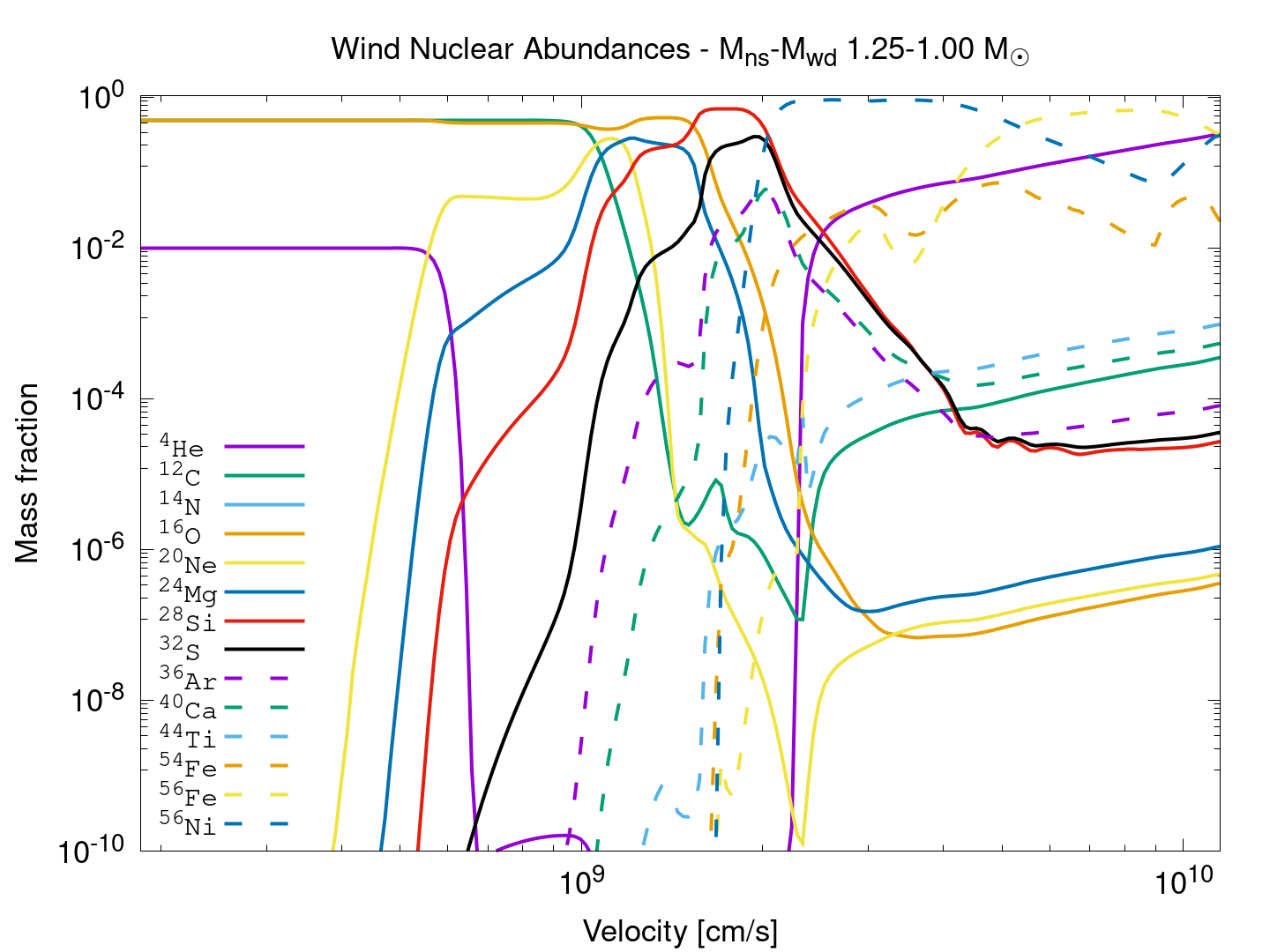}
  \includegraphics[width=.45\linewidth]{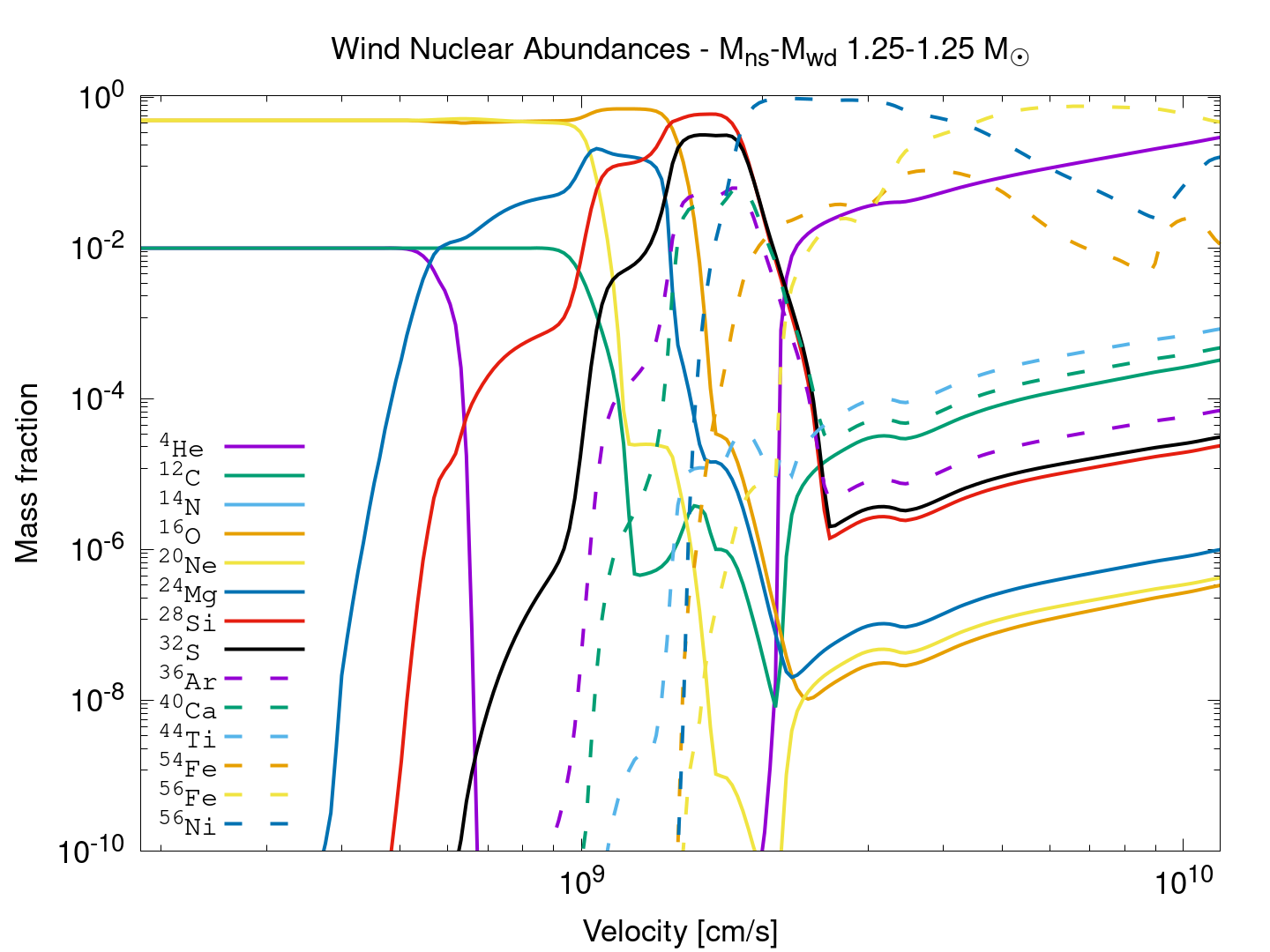}
  \caption{(Top Left) Initial WD mass $0.50 M_\odot$. (Top Right) Initial WD mass $0.75 M_\odot$. (Bottom Left) Initial WD mass $1.00 M_\odot$. (Bottom Right) Initial WD mass $1.25 M_\odot$. All companion NSs are $1.25 M_\odot$. Here we have the wind-ejecta nuclear abundances. The abundances are grouped by ejection velocity and summed over the entirety of the simulation, about 60s. When looking at the distribution of isotopes in the wind, we find that the fastest moving material in the ejecta is composed primarily of $^{4}$He, $^{54,56}$Fe, and $^{56}$Ni.}
  \label{fig:ewb}
\end{figure}

\begin{figure}[h]
    \centering
    \includegraphics[width=0.45\columnwidth]{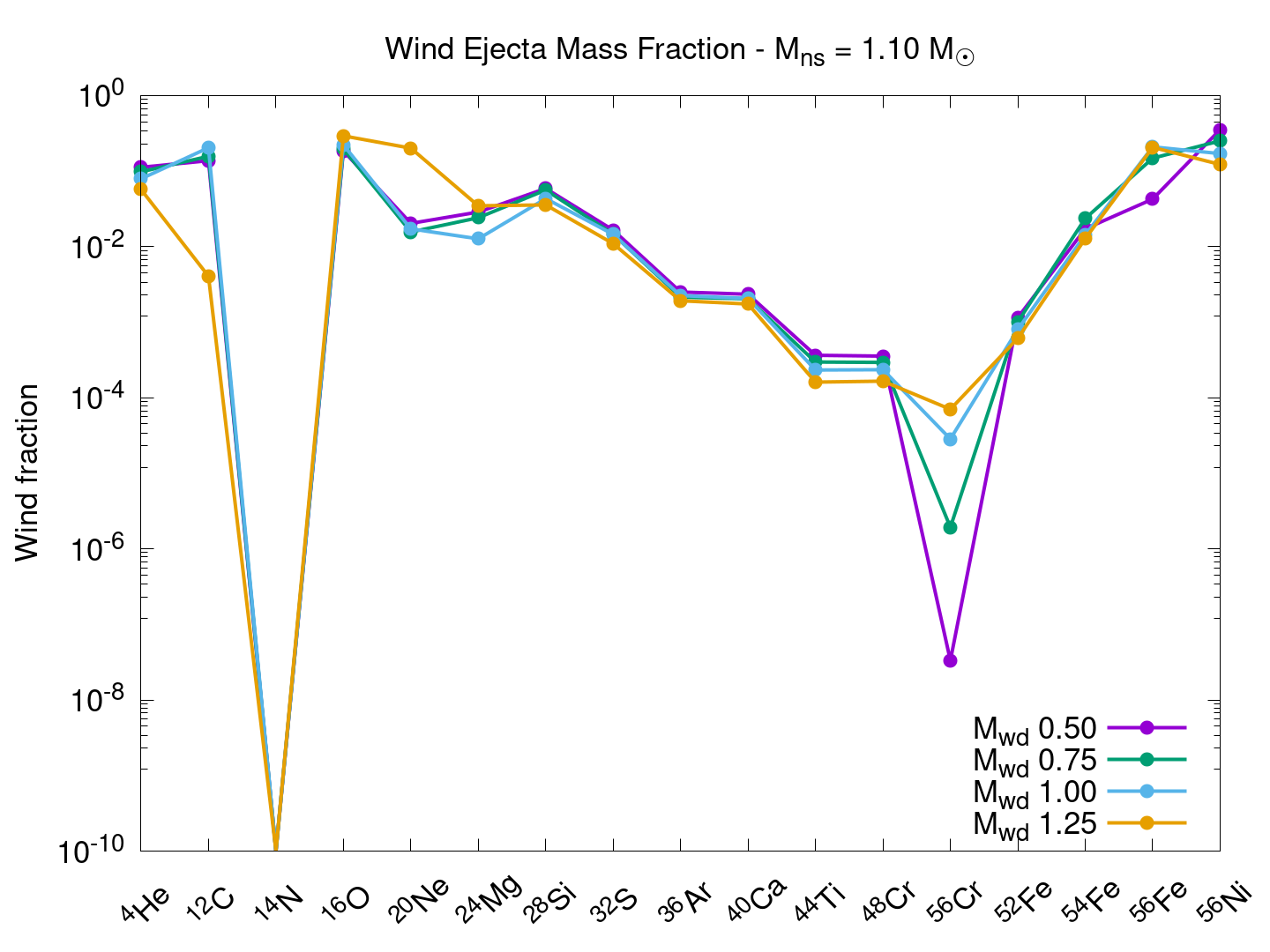}
    \includegraphics[width=0.45\columnwidth]{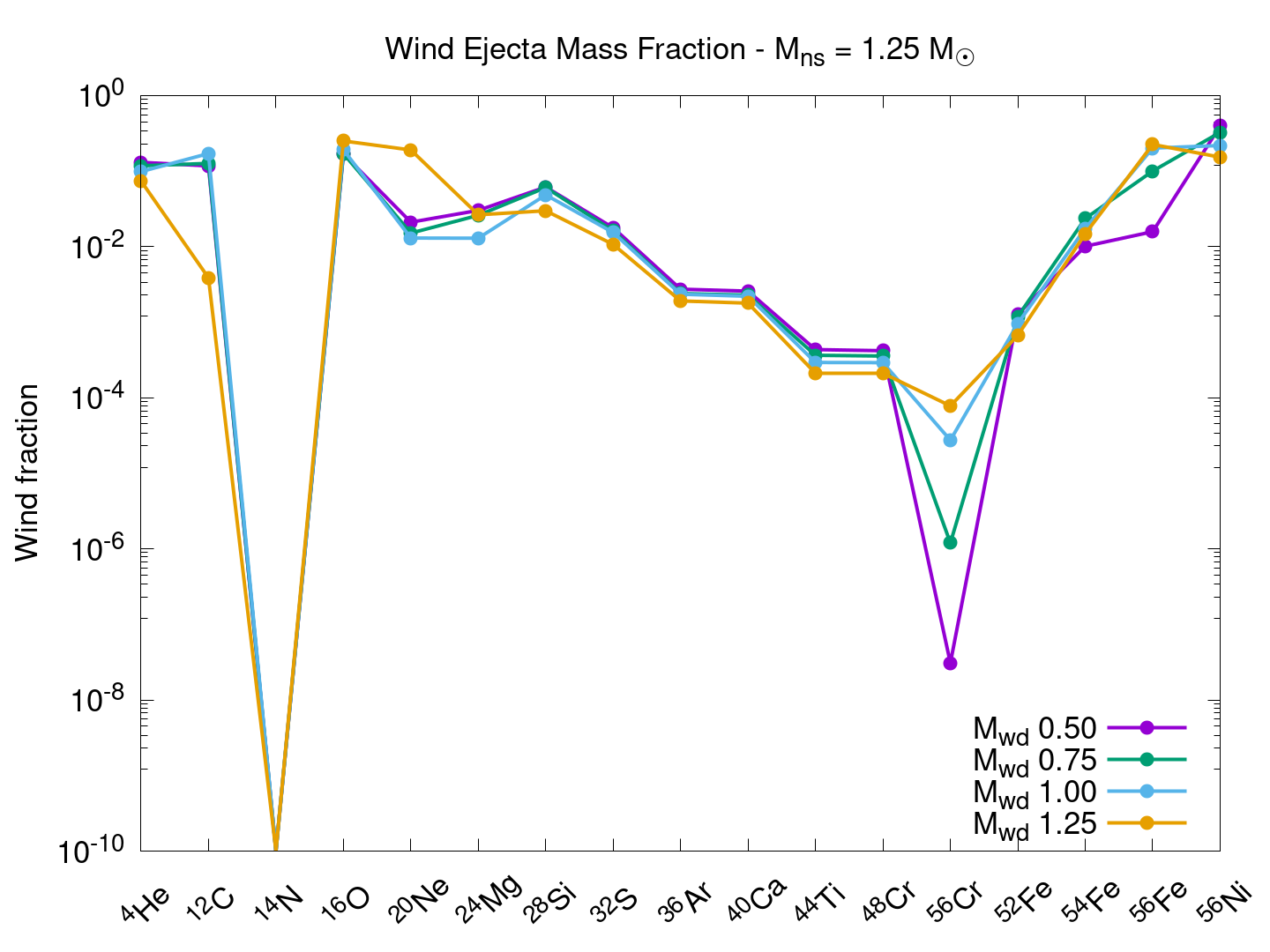}\\
    \includegraphics[width=0.45\columnwidth]{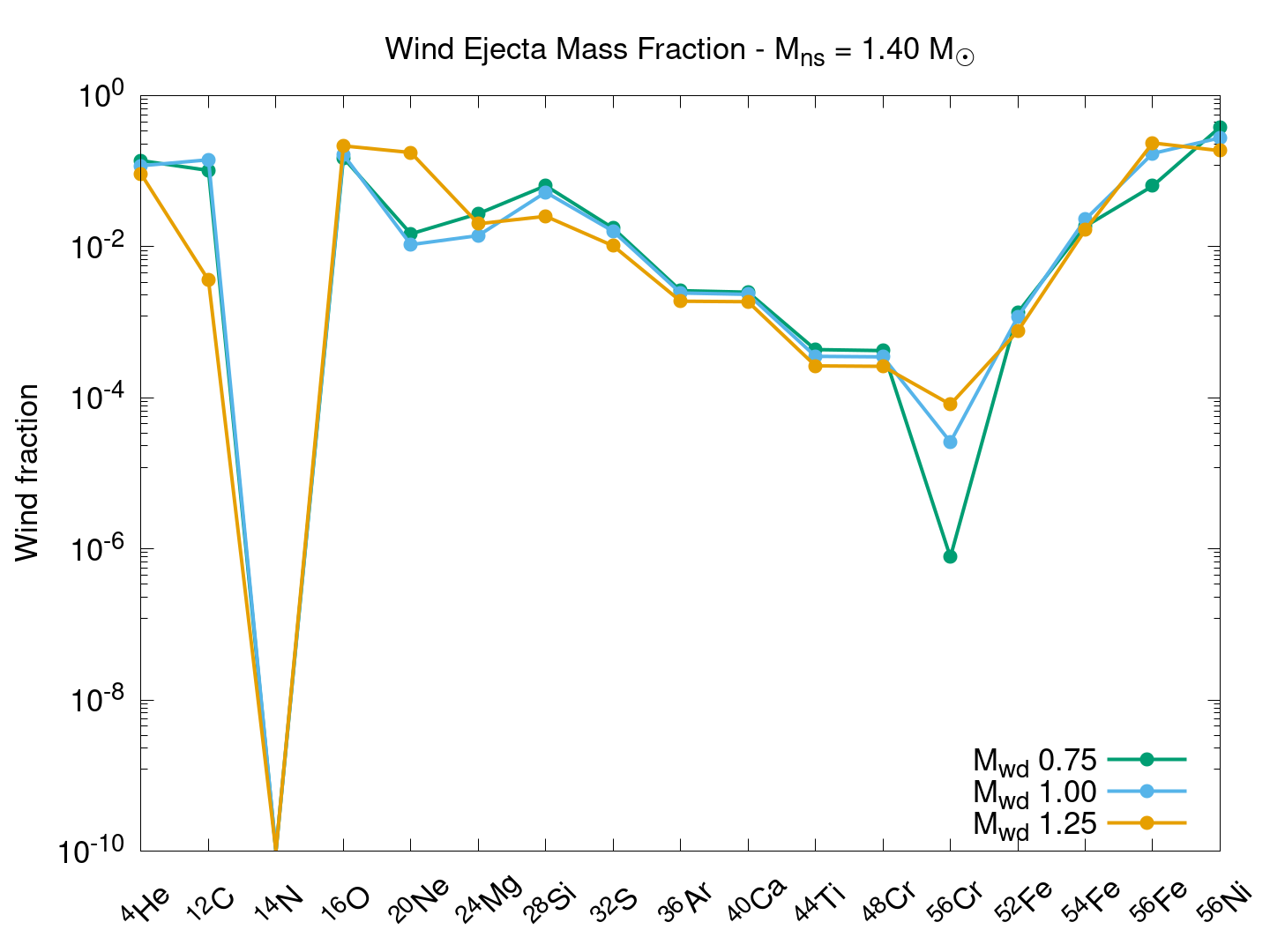}
    \includegraphics[width=0.45\columnwidth]{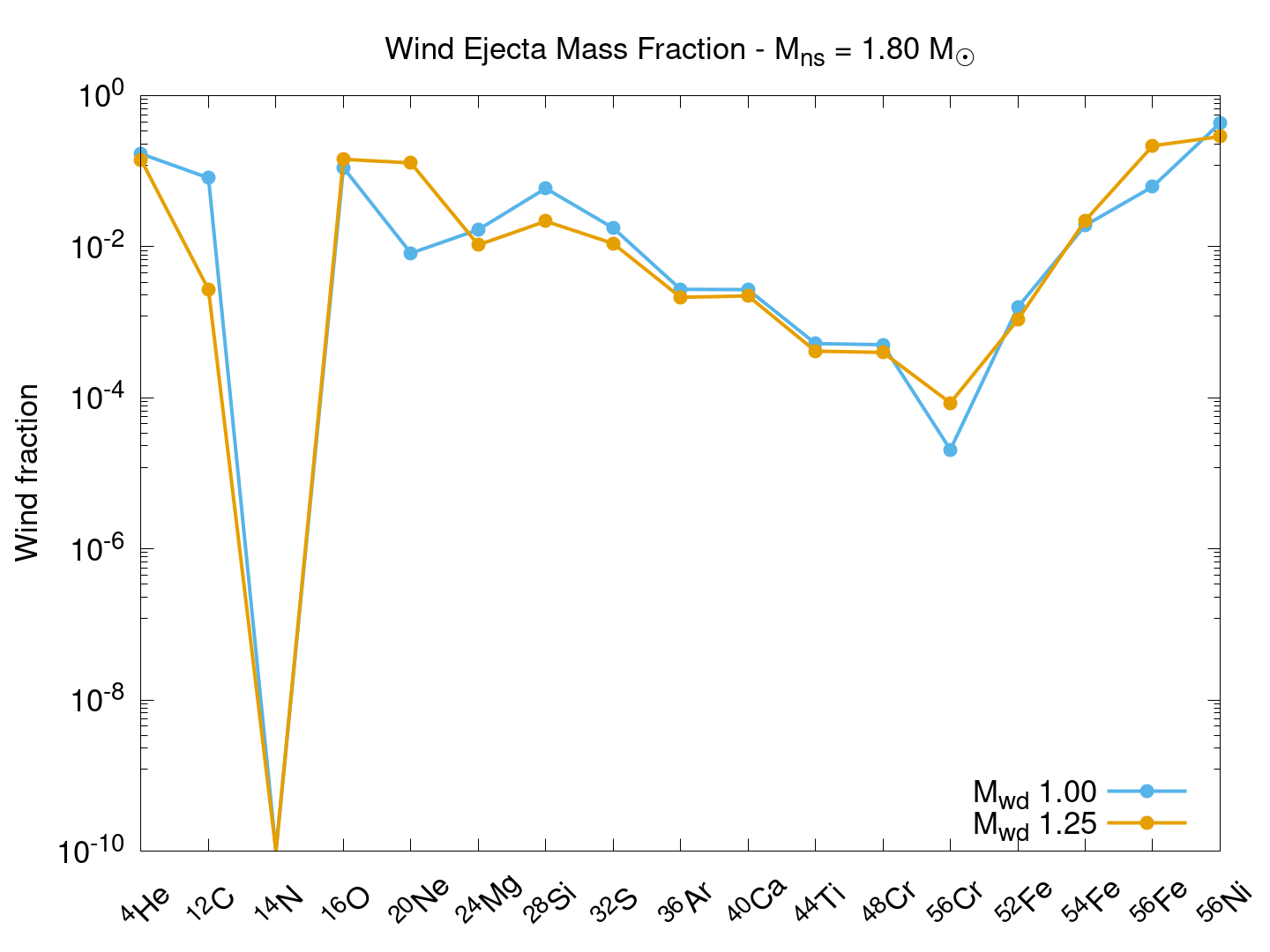}
    \caption{We have plotted the wind abundances for each of the configurations. (Top Left) Initial NS mass $1.10 M_\odot$. (Top Right) Initial NS mass $1.25 M_\odot$. (Bottom Left) Initial NS mass $1.40 M_\odot$. (Bottom Right) Initial NS mass $1.80 M_\odot$. Using the 21-isotope nuclear network, we highlight a subset of the available isotopes. Of particular interest are the abundances of $^{44}$Ti, $^{56}$Fe, and $^{56}$Ni found in the ejecta.}
    \label{fig:windabundance}
\end{figure}

The disk-wind abundances are determined by adiabatically expanding the ejected material at each timestep. The ejected material over the entire simulation is binned by velocity to get the bulk wind abundances, see Figure~\ref{fig:ewb}. First, we used the 21-isotope network. Of particular note is the fastest moving material, i.e., the material ejected from the inner most regions of the disk $r<1\times 10^7$ cm. This material largely comprises $^{54}$Fe, $^{56}$Fe, and $^{56}$Ni. We find a significant fraction of $^{44}$Ti in these disk winds (see Figure~\ref{fig:windabundance}).

\begin{table}[h]
    \begin{tabular}{|l|l|l|l|l|c|c|c|}
    \hline
    ~ & 0.50 $M_{\odot}$ & 0.75 $M_{\odot}$ & 1.00 $M_{\odot}$ & 1.25 $M_{\odot}$ & 15 $M_{\odot}$ & 20 $M_{\odot}$ & 25 $M_{\odot}$\\ \hline
    $^{26}$Al & 3.78E-07 & 4.60E-07 & 2.53E-07 & 1.73E-06 & 1.79E-03 & 1.29E-09 & 1.62E-08 \\ \hline
    $^{27}$Al & 5.09E-06 & 5.84E-06 & 3.09E-06 & 1.61E-05 & 1.01E-04 & 7.37E-08 & 2.95E-07 \\ \hline
    $^{40}$Ca & 7.60E-05 & 1.12E-04 & 1.89E-04 & 1.84E-04 & 4.77E-03 & 1.79E-04 & 9.91E-03 \\ \hline
    $^{48}$Ca & 9.96E-07 & 8.93E-04 & 1.68E-03 & 4.31E-04 & 3.39E-05 & 3.02E-04 & 8.96E-04 \\ \hline
    $^{44}$Ti & 3.13E-07 & 4.04E-07 & 5.17E-07 & 3.94E-07 & 1.78E-03 & 7.08E-03 & 1.11E-02 \\ \hline
    $^{50}$Ti & 2.31E-04 & 9.54E-04 & 1.21E-03 & 1.22E-03 & 1.62E-04 & 4.48E-04 & 3.81E-04 \\ \hline
    $^{52}$Fe & 2.51E-05 & 4.13E-05 & 6.04E-05 & 5.84E-05 & 1.79E-03 & 1.29E-09 & 1.62E-08 \\ \hline
    $^{53}$Fe & 4.51E-06 & 4.41E-06 & 4.24E-06 & 4.50E-06 & 1.01E-04 & 7.37E-08 & 2.95E-07 \\ \hline
    $^{54}$Fe & 1.27E-04 & 4.32E-04 & 7.44E-04 & 1.99E-03 & 4.77E-03 & 1.79E-04 & 9.91E-03 \\ \hline
    $^{55}$Fe & 7.57E-06 & 4.22E-05 & 6.46E-05 & 1.71E-04 & 3.39E-05 & 3.02E-04 & 8.96E-04 \\ \hline
    $^{56}$Fe & 1.34E-03 & 2.11E-03 & 2.66E-03 & 5.27E-03 & 1.78E-03 & 7.08E-03 & 1.11E-02 \\ \hline
    $^{57}$Fe & 6.24E-05 & 7.40E-05 & 7.45E-05 & 1.33E-04 & 1.62E-04 & 4.48E-04 & 3.81E-04 \\ \hline
    $^{58}$Fe & 2.23E-03 & 2.76E-03 & 2.71E-03 & 4.14E-03 & 3.66E-04 & 1.36E-03 & 1.43E-03 \\ \hline
    $^{59}$Fe & 6.56E-06 & 2.03E-05 & 2.73E-05 & 3.18E-05 & 5.48E-05 & 3.89E-05 & 1.13E-04 \\ \hline
    $^{60}$Fe & 7.92E-05 & 9.62E-04 & 1.58E-03 & 1.59E-03 & 3.19E-05 & 2.13E-05 & 9.80E-05 \\ \hline
    $^{61}$Fe & 5.01E-09 & 6.76E-07 & 1.84E-06 & 6.74E-07 & 4.23E-07 & 6.91E-10 & 3.43E-06 \\ \hline
    $^{62}$Fe & 1.24E-08 & 4.02E-05 & 7.79E-05 & 1.49E-05 & 9.76E-09 & 1.65E-13 & 1.29E-07 \\ \hline
    $^{64}$Fe & 1.64E-15 & 1.33E-07 & 1.90E-06 & 5.94E-10 & 3.10E-28 & 7.94E-29 & 4.78E-25 \\ \hline
    $^{66}$Fe & 1.31E-23 & 1.57E-09 & 1.72E-05 & 2.03E-15 & 4.18E-29 & 7.07E-31 & 1.50E-28 \\ \hline
    $^{56}$Ni & 6.65E-03 & 8.54E-03 & 9.95E-03 & 1.32E-02 & 9.18E-02 & 3.52E-09 & 1.12E-07 \\ \hline
    $^{57}$Ni & 2.04E-04 & 3.27E-04 & 4.93E-04 & 6.88E-04 & 3.75E-03 & 1.14E-07 & 2.73E-07 \\ \hline
    $^{58}$Ni & 2.21E-03 & 3.06E-03 & 3.95E-03 & 6.24E-03 & 5.18E-02 & 8.71E-05 & 2.76E-04 \\ \hline
    $^{59}$Ni & 4.71E-05 & 6.08E-05 & 7.81E-05 & 1.06E-04 & 4.37E-04 & 2.07E-05 & 2.73E-05 \\ \hline
    $^{60}$Ni & 1.41E-03 & 1.63E-03 & 1.88E-03 & 2.03E-03 & 3.63E-03 & 7.98E-04 & 2.10E-03 \\ \hline
    $^{61}$Ni & 1.85E-05 & 1.89E-05 & 1.92E-05 & 1.90E-05 & 7.41E-05 & 1.71E-04 & 2.31E-04 \\ \hline
    $^{62}$Ni & 1.08E-03 & 1.10E-03 & 1.02E-03 & 1.33E-03 & 1.07E-03 & 7.74E-04 & 2.43E-03 \\ \hline
    $^{63}$Ni & 3.62E-06 & 5.66E-06 & 6.54E-06 & 8.43E-06 & 3.63E-05 & 9.90E-05 & 1.61E-04 \\ \hline
    $^{64}$Ni & 3.28E-04 & 1.11E-03 & 1.70E-03 & 2.15E-03 & 9.31E-05 & 2.46E-04 & 6.40E-04 \\ \hline
    $^{65}$Ni & 5.33E-07 & 9.41E-06 & 1.83E-05 & 1.64E-05 & 2.61E-06 & 4.83E-08 & 1.95E-05 \\ \hline
    $^{66}$Ni & 1.71E-05 & 2.04E-03 & 4.12E-03 & 3.01E-03 & 6.42E-07 & 1.34E-08 & 6.54E-06 \\ \hline
    $^{67}$Ni & 1.92E-09 & 7.18E-06 & 5.13E-05 & 3.42E-06 & 2.10E-11 & 2.78E-16 & 5.61E-10 \\ \hline
    $^{68}$Ni & 3.05E-09 & 5.23E-04 & 1.88E-03 & 5.97E-05 & 1.11E-12 & 6.22E-18 & 6.45E-11 \\ \hline
    \end{tabular}
    \caption{This table collects a selection of isotopes, with attention to heavy Fe and Ni by post-processing nuclear yields using the 495-isotope Torch network for the efficient-wind results. The WD masses are labeled at the top of columns 2-5, each with a $1.25 M_\odot$ NS companion. Columns 6,7, and 8 are median nuclear yields from many core-collapse supernovae simulations \citep[][]{Andrews2020}. The mass in the header corresponds to the initial mass of the single star before collapse. The ion name is listed in the leftmost column, and the data are given in solar mass, $M_\odot$.}
    \label{tab:heavyFeNi}
\end{table}

\begin{figure}[h]
    \centering
    \includegraphics[width=0.45\textwidth]{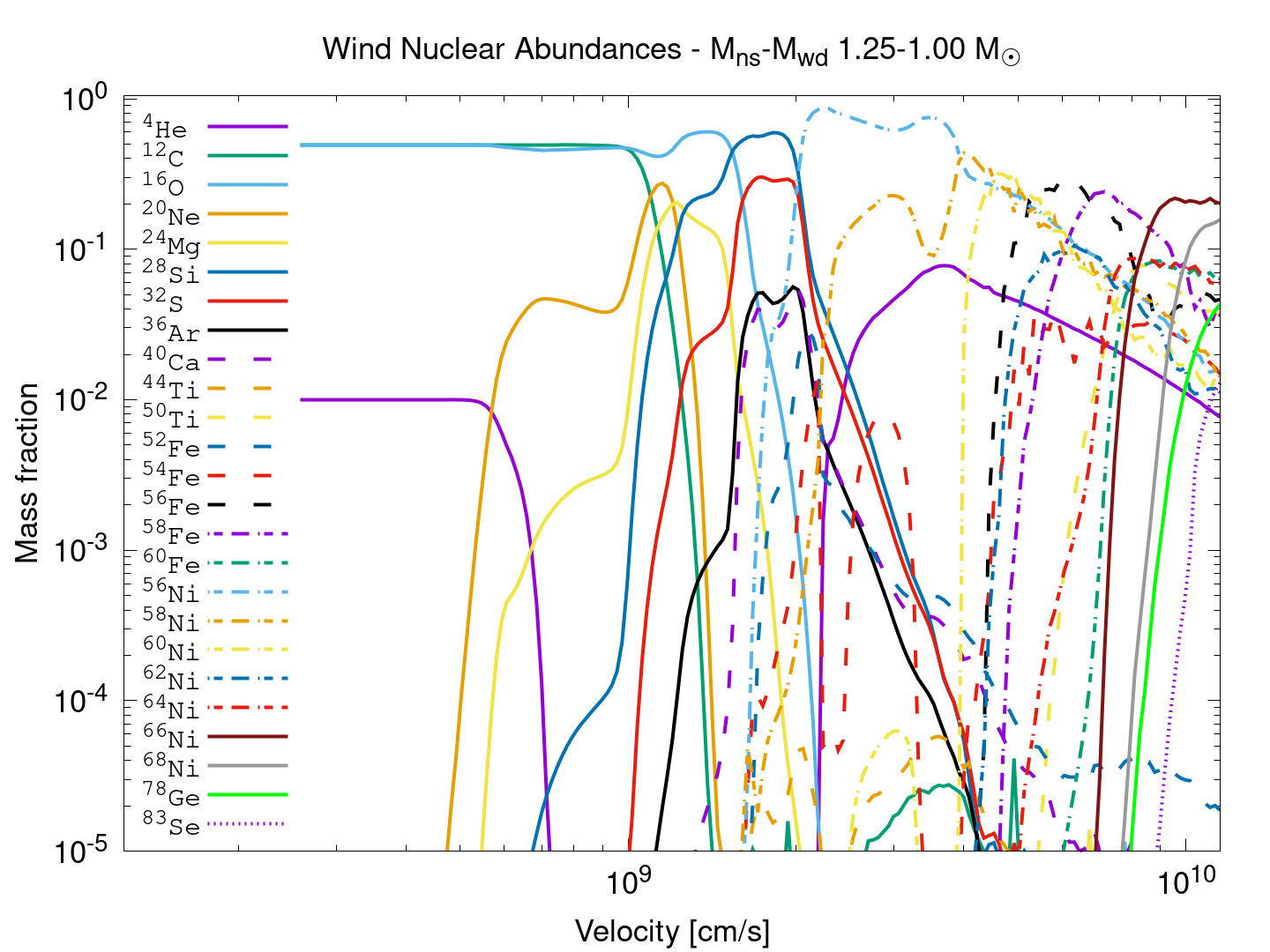}
    \includegraphics[width=0.45\textwidth]{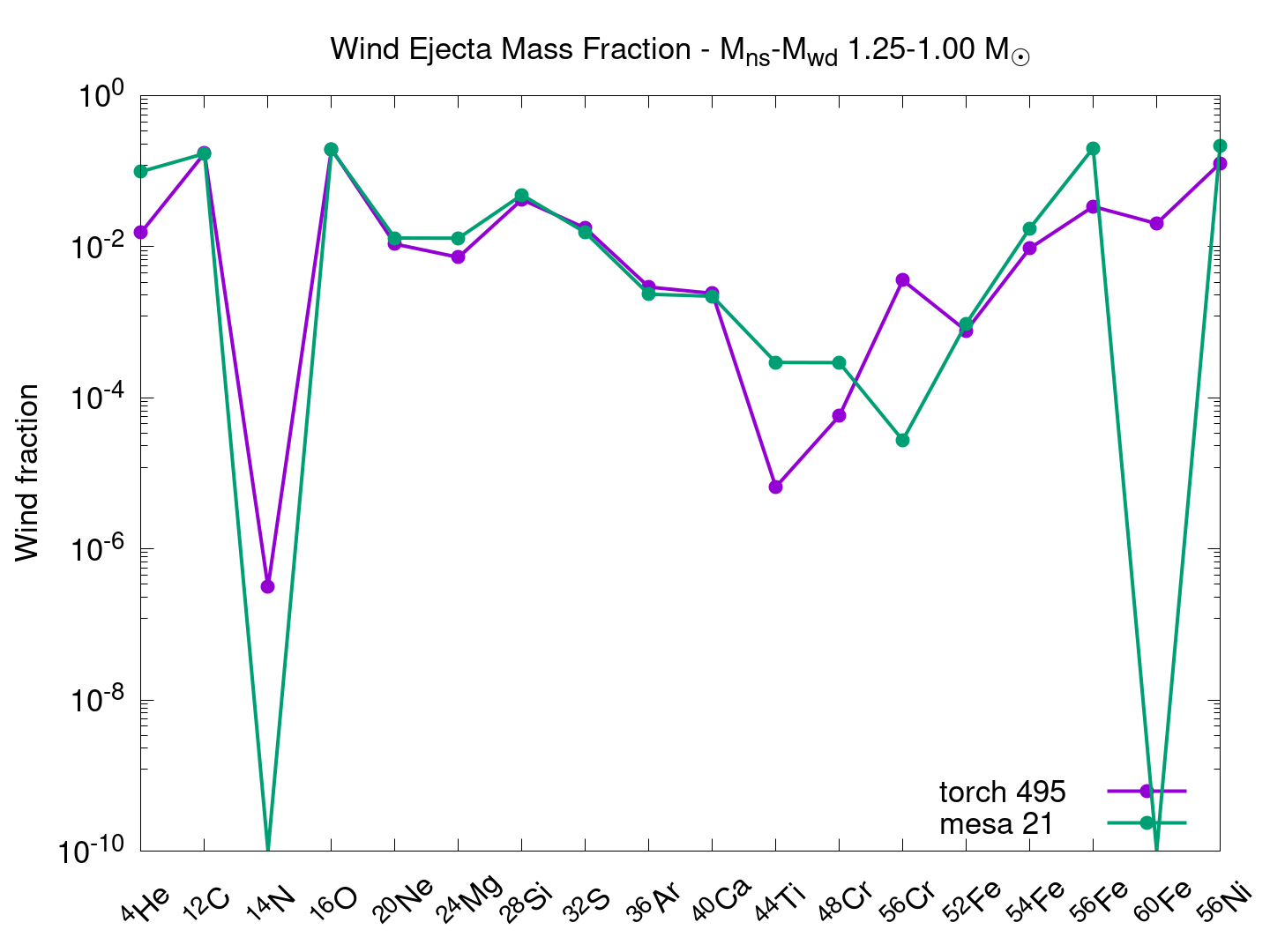}
    \caption{(Left) The wind-ejecta nuclear abundances for NS mass $1.25 M_\odot$ and WD mass $1.00 M_\odot$. Compared with the simpler network, we find that the diversity of isotopes moving faster than $2\times10^9$\,cm/s increased distinctly. The fastest moving material is comprised largely of heavy iron (e.g., $^{60}$Fe), and heavy nickel (e.g., $^{66}$Ni). (Right) A reduced subset of the nuclear abundances are compared to the simpler 21-isotope network. The $^{44}$Ti became fuel for the production of other isotopes, being reduced by nearly two orders of magnitude.}
    \label{fig:torchbundance}
\end{figure}
$^{44}$Ti is known to be burned into a myriad of heavier elements, and our resulting abundances show large amounts of this isotope. In order to investigate outflows in more depth, we took the wind results and processed them using the 495-isotope Torch network. Select results are shown in Figure~\ref{fig:torchbundance}. It comes as little surprise that the abundances of $^{44}$Ti decrease substantially, and the mass fractions of elements are distributed over more isotopes. The production of significant fractions of heavier Fe and Ni are of special note. Due to the high-accretion rates found in the disks, we see that the innermost regions of the disks reach densities and temperatures that promote deleptonization of the material through electron-capture. While the rates are not massive, they are just enough to produce neutron rich Fe and Ni isotopes in significant quantities. Coupling the inner burning regions with the wind-prescription, we find that the fastest-moving material ejected in the wind, i.e., the front of the wind, largely comprises these neutron-rich, iron-peak isotopes, particularly $^{60}$Fe.

$^{60}$Fe is a long-lived radioactive isotope with a half-life of $2.62 \times 10^6$ yr. The signatures of this isotope have been detected from the decay lines measured in the diffuse galactic gamma-ray foreground of the Milky Way, as well as the gamma-ray fluxes from the interstellar medium \citep{Bishop2011,Bouchet2011,Mahoney1982}. Not only has $^{60}$Fe been detected in space, but also terrestrial deposits have been detected, suggesting a more recent accumulation \citep{Gounelle2012,Wallner2016}. Since this isotope is not naturally occurring, it must have accreted on Earth from outside the solar system. Candidates for $^{60}$Fe production have been suggested from core-collapse supernovae \citep{Jones2019}, electron-capture supernovae \citep{Wanajo2013}, high-density type-Ia supernovae \citep{Woosley1997}, and super-AGB stars \citep{Lugaro2012}. The results of the 495-Torch analysis suggest $^{60}$Fe is a fast-moving, significant fraction of the wind across all models, and, thus, NS-WD mergers may help explain these observations and Galactic chemical distributions.

Table~\ref{tab:heavyFeNi} collects a selection of isotopes, specifically the heavy Fe and Ni, found in the wind of these mergers and compares them with results from CCSNe simulations of three progenitor masses, 15 $M_{\odot}$, 20 $M_{\odot}$, and 25 $M_{\odot}$. The neutron-rich isotopes are the most interesting aspects of the NS-WD yields. In particular, where yields from the mergers are $\sim10^2$ larger than the CCSNe, the yields contribute to Galactic distributions. This is due to the rate of these mergers being 100 times less frequent than SNe. There are a few isotopes that are produced in significantly larger quantities (10$^3$ times greater) than that of the CCSNe, and thus the yields from mergers could dominate the yield distribution. $^{60}$Fe, $^{64}$Fe, $^{66}$Fe, $^{64}$Ni, $^{66}$Ni, $^{67}$Ni, and $^{68}$Ni both are produced in large enough quantities as well as being larger than the CCSNe yields. We already have discussed the interest in $^{60}$Fe. The heavier Fe isotopes are unstable with half-lives on the order of seconds. These heavy Fe nuclei decay relatively quickly towards stable $^{64}$Ni. $^{66}$Ni, $^{67}$Ni, and $^{68}$Ni decay with half-lives from seconds to hours to stable $^{66}$Zn, $^{67}$Zn, and $^{68}$Zn.

$^{64}$Ni is produced in large quantities from these mergers; however, the yields are only on the order of 10 times larger than CCSNe yields. NS-WD mergers can contribute to $^{64}$Ni, but these events do not dominate the distribution. $^{60}$Fe remains the unique neutron-rich isotope produced from these mergers.

Another unique aspect of these mergers is the production of some of the heavier isotopes. As seen in Figure \ref{fig:torchbundance}, the mergers produce a significant fraction of $^{78}$Ge and $^{83}$Se. The heavier the initial WD mass the more isotopes heavier than Ni are produced. The 0.50 $M_\odot$ initial mass begins producing $^{59,61,67,68,69}$Cu, $^{62,72}$Zn, $^{75}$Ga, $^{78}$Ge, and $^{82,83}$Se. The 0.75 $M_\odot$ additionally burns up to $^{67}$Ni, $^{67}$Cu, and $^{72}$Zn, while the 1.00 $M_\odot$ also produces $^{66}$Fe, $^{66,67}$Co, and $^{67}$Cu. The 1.25 $M_\odot$ was initially an ONe star; many of these isotopes are present but not as prominent as the initial CO stars.

The neutron-rich, iron-peak elements are predominantly in the outer fastest moving ejecta, such that when they decay, the radiation is able to escape and stream freely away after the earliest times. A few of these isotopes decay with energies and half-lives comparable to $^{56}$Ni (6.077d). The $^{59}$Ni decay emission is of the same order with a longer half-life (44.503d). The rest would be relevant for the first couple of days after the merger. Other relevant emissions come from $^{57}$Ni, $^{66}$Ni, $^{66}$Cu, $^{67}$Cu, and $^{72}$Zn, with energies and half-lives summarized in Table \ref{tab:decay}. These neutron-rich, iron-peak elements will produce distinct gamma-ray decay emissions. 

\begin{table}[h]
\centering
    \begin{tabular}{|c|c|c|c|}
    \hline
     & Q-value (keV) & Half-life & $M_{ej} $($M_\odot$)\\
    \hline
    $^{56}$Ni & 2135.0 & 6.077d & 9.95E-03\\
    \hline
    $^{59}$Fe & 1565.2 & 44.503d & 2.73E-05 \\
    \hline
    $^{57}$Ni & 3264.0 & 35.60h & 4.93E-04 \\
    \hline
    $^{66}$Ni & 226.00 & 54.60h & 4.12E-03 \\
    \hline
    $^{66}$Cu & 2642.0 & 5.120m & 6.59E-08* \\
    \hline
    $^{67}$Cu & 577.00 & 61.830h & 1.55E-05 \\
    \hline
    $^{72}$Zn & 458.00 & 46.50h & 4.17E-05 \\
    \hline
    \end{tabular}
    \caption{In this table, we provide isotopes with decay properties relevant for the timing of the wind ejecta for the NS-WD 1.25-1.00 merger. The isotopes collected here all decay within the first few days of the merger with energies and mass fractions significant enough to contribute to the spectra. The initial wind ejecta mass of $^{66}$Cu is quite low in comparison with the other isotopes; however, it is the dominant decay product of $^{66}$Ni, and thus $^{66}$Cu will be produced and decay on a similar time scale as $^{66}$Ni.}
    \label{tab:decay}
\end{table}

\section{Observable Transients}\label{sec:observables}
\subsection{Light Curves and Spectra Properties}
\label{sec:snlc}

The disk ejecta from the wind can produce a sub-luminous, hydrogen-poor supernova-like transient. We considered three sets of models as described in Section~\ref{sec:numresults}: high-entropy, low-entropy, and low-entropy efficient-wind models. Our disk models calculate the composition and mass ejected at every point in the disk. We derive the velocity of this ejecta by assuming it is roughly the escape velocity of the material at the disk based on recent calculations of disk winds, albeit from neutron star merger disks~\citep[e.g.][]{2019PhRvD.100b3008M}. The inner material that undergoes the most significant nuclear processing is ejected at the highest velocity (see Figure~\ref{fig:mlb}). These high velocities mean that this material leads the wind shock and becomes optically thin quickly. Even though the $^{56}$Ni yield is lower than most supernovae, most of the gamma-rays produced in its decay escape. From the gamma-ray flux of these outflows we expect gamma-ray luminosities peaking in the range of $10^{39}-10^{42} {\rm \, erg \, s^{-1}}$.

These initial ejecta properties are input into the {\it SuperNu} light-curve code~\citep{2013ApJS..209...36W} that has been used for a variety of astrophysical transients including type Ia supernovae~\citep{2016ApJ...827..128V}, type II supernovae~\citep{2017ApJ...845..168W}, and kilonovae~\citep{2018MNRAS.478.3298W}. It models the transport of radioactive gamma-ray lines using a single-group opacity to capture this radioactive heating and models the multi-group UVOIR spectra and light-curve emission. For our calculations, we discretize the photon energy into 500-groups, providing rough spectra and well-sampled light-curve bands. Although {\it SuperNu} is capable of multi-dimensional models, for this study we focus on spherically-symmetric explosions.

\begin{figure}[h]
    \includegraphics[width=0.33\textwidth]{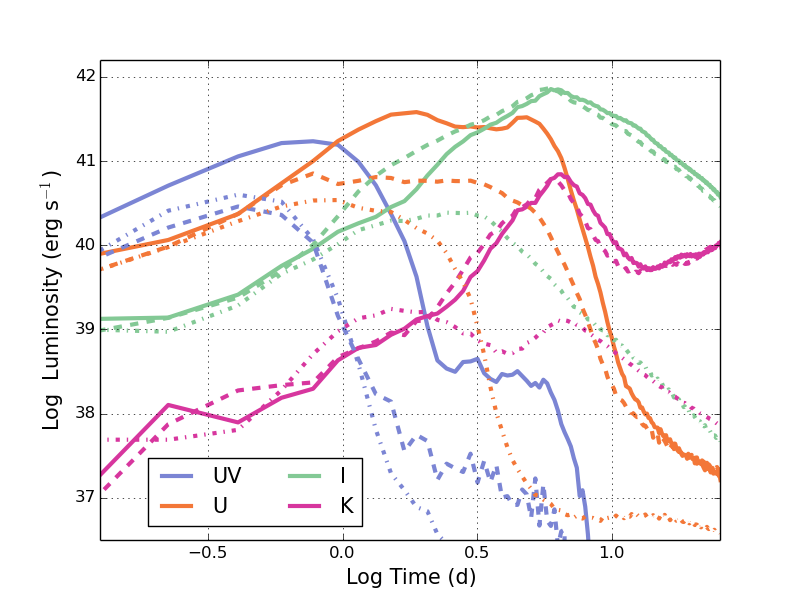}
    \includegraphics[width=0.33\textwidth]{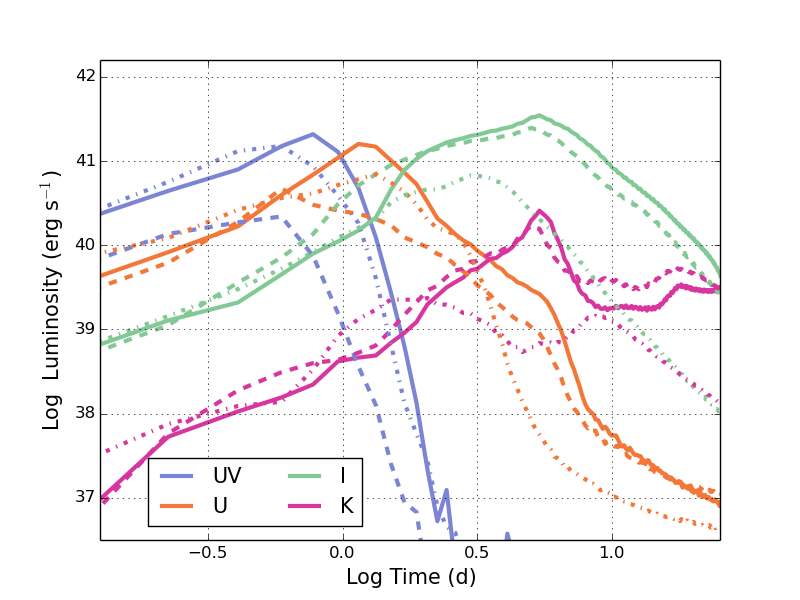}
    \includegraphics[width=0.33\textwidth]{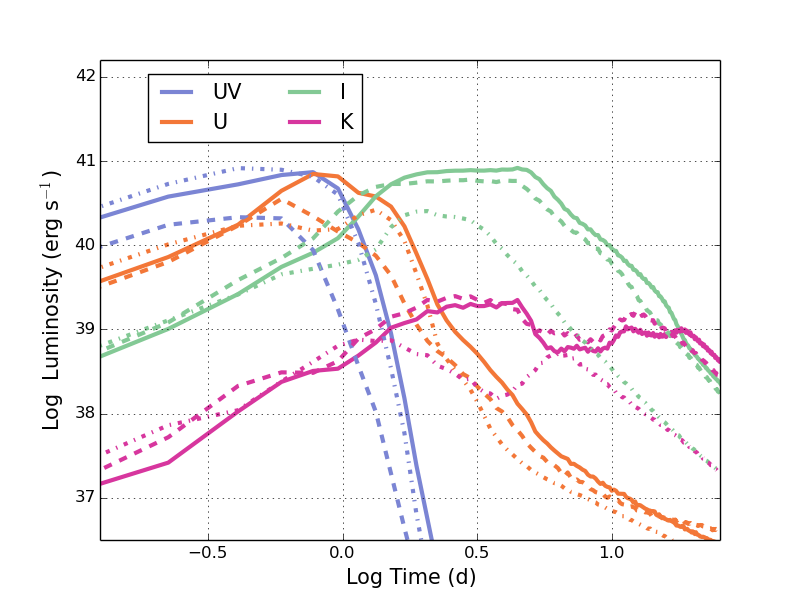}
    \caption{Luminosity as a function of time in the UV-, U-, I-, and K-bands for our three different disk models: high-entropy (left), low-entropy (middle), and efficient-wind (right). The linestyles correspond to different mass pairings of the NS and WD: 1.8\,M$_\odot$ NS, 1.25\,M$_\odot$ WD (solid), 1.1\,M$_\odot$ NS, 1.25\,M$_\odot$ WD (dashed), and 1.1\,M$_\odot$ NS, 0.5\,M$_\odot$ WD (dash-dot). All models produce similar trends with a strong UV component in the first day and a peak in the I-band at 5-8d. The models with more ejecta mass (high-entropy case) produce the most luminous light-curves.}
    \label{fig:lcmodels}
\end{figure}

The luminosity in UV-, U-, I-, and K-bands for three different NS-WD mass pairings and our three disk models is shown in Figure~\ref{fig:lcmodels}. All of these models produce similar trends with a strong UV emission (exceeding $10^{41} {\rm \, erg \, s^{-1}}$) in the first day and peaking in the I-band (as well as the bolometric) at 5-8d with luminosities lying between $10^{41}-10^{42} {\rm \, erg \, s^{-1}}$. The models with higher ejecta mass (high-entropy models) produce the most luminous light-curves. The early-time UV emission is driven by the fact that the radioactive isotopes are located in the outermost ejecta layers. This outward distribution of radioactive isotopes, combined with the low ejecta-mass and high velocities (relative to core-collapse supernovae) means that these light-curves peak early ($<10d$) and decline quickly.
   
\begin{figure}[h]
    \includegraphics[width=0.45\textwidth]{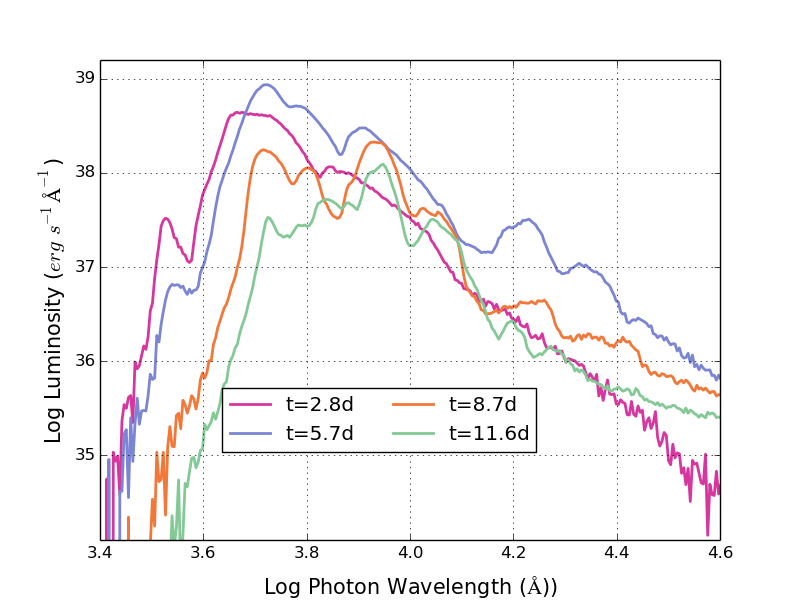}
    \includegraphics[width=0.45\textwidth]{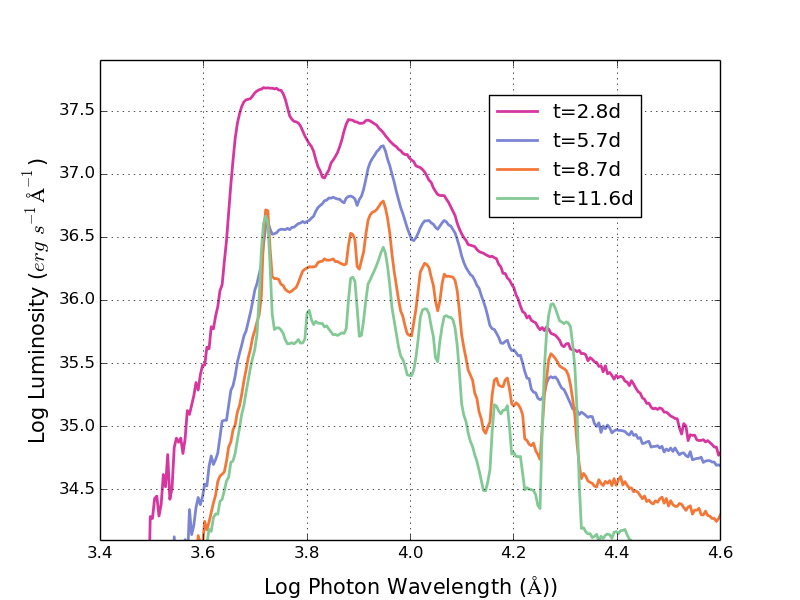}
    \caption{Emission versus wavelength for the high-entropy (left) and efficient-wind (right) models with a $1.25\,M_\odot$ WD being disrupted by a $1.1\,M_\odot$ NS.}
    \label{fig:specmodels}
\end{figure}

Due to the high velocities in our ejecta (because we set the velocities to the escape velocities, the ejecta velocity distribution ranges from $1-100\, {\rm km \, s^{-1}}$) the spectra of these transients include a set of broad line features. Figure~\ref{fig:specmodels} shows the spectra for our high-entropy and efficient-wind models with a $1.1\,M_\odot$ NS and a $1.25\,M_\odot$ WD.

\begin{figure}[h]
    \includegraphics[width=0.45\textwidth]{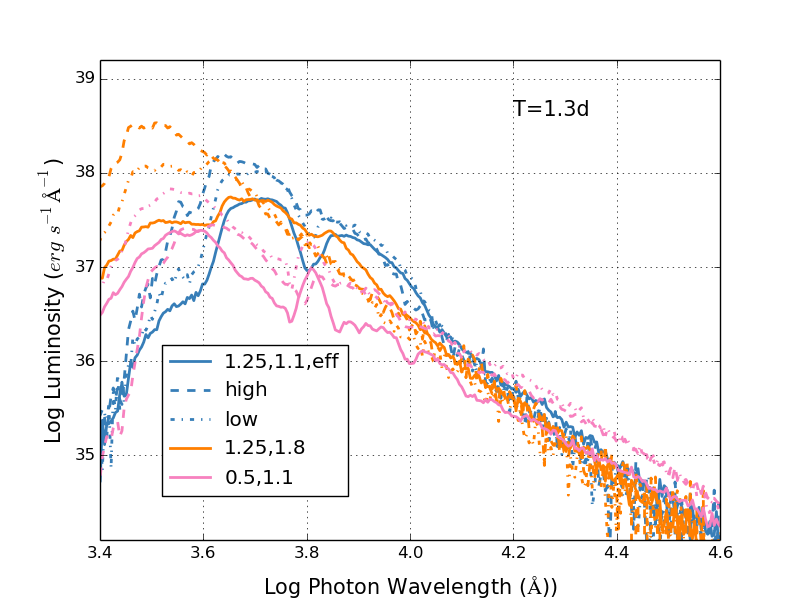}
    \includegraphics[width=0.45\textwidth]{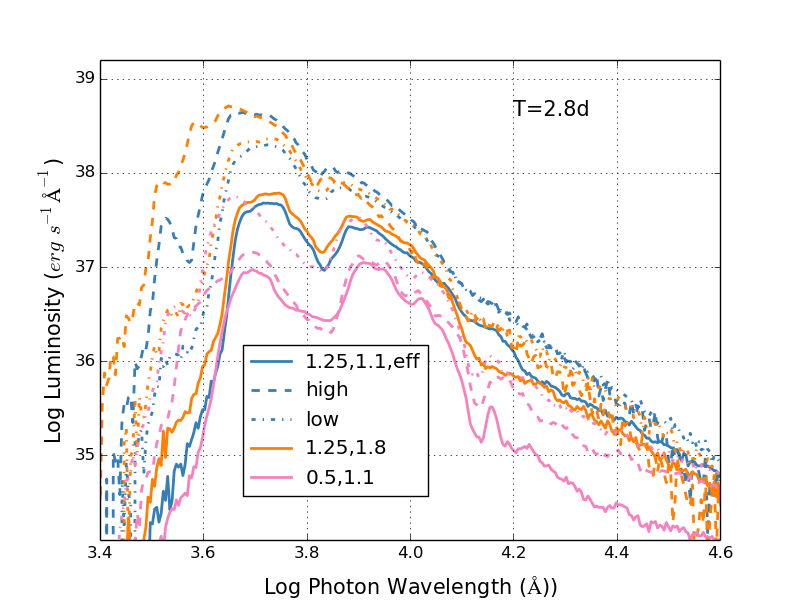}
    \includegraphics[width=0.45\textwidth]{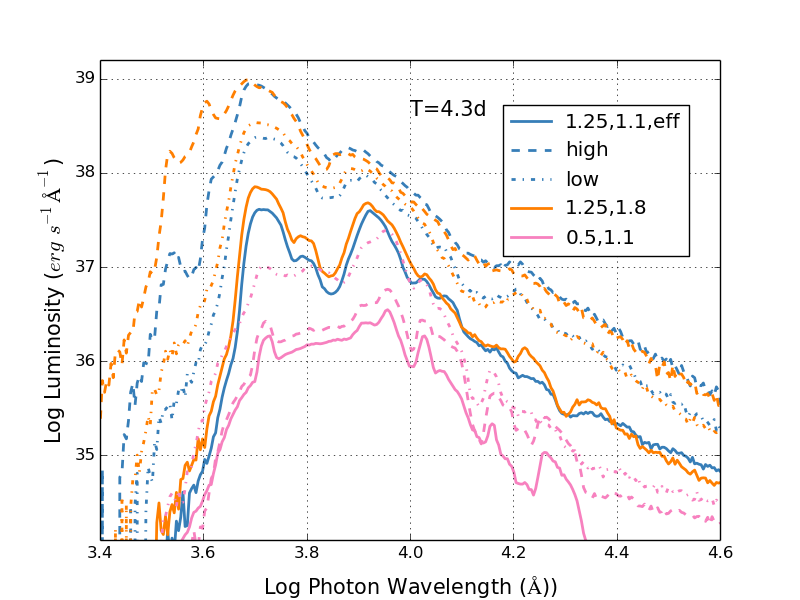}
    \caption{Spectra for a range of models varying the WD and NS mass as well as the disk model: high entropy, low entropy, and efficient wind. We follow these models at three early timescales at 1.3, 2.8, 4.3d. These models show the range of spectra from our standard ejecta-velocity models.}
    \label{fig:specat}
\end{figure}

Transient AT2018kzr may well be an example of such WD mergers~\citep{2020MNRAS.497..246G}. With its broad velocities and rapid evolution, AT2018kzr exhibits many of the features seen in models of WD mergers. Figure~\ref{fig:specat} show spectra for a range of our NS-WD merger models. These spectra produce a range of models with a range of spectral properties. From our models, the peak wavelength of the emission lies somewhere between 2500-5000$\AA$ at 1.3\,d and 5000-9000$\AA$ at 4.3\,d. These models encompass the spectra observed for AT2018kzr. For our models, the features are quite broad, but the velocities could be lower, producing narrower line features and slower transient evolution. We confirm the conclusions from \cite{2020MNRAS.497..246G} that this event could be a NS-WD merger.

\begin{figure}[h]
    \includegraphics[width=0.45\textwidth]{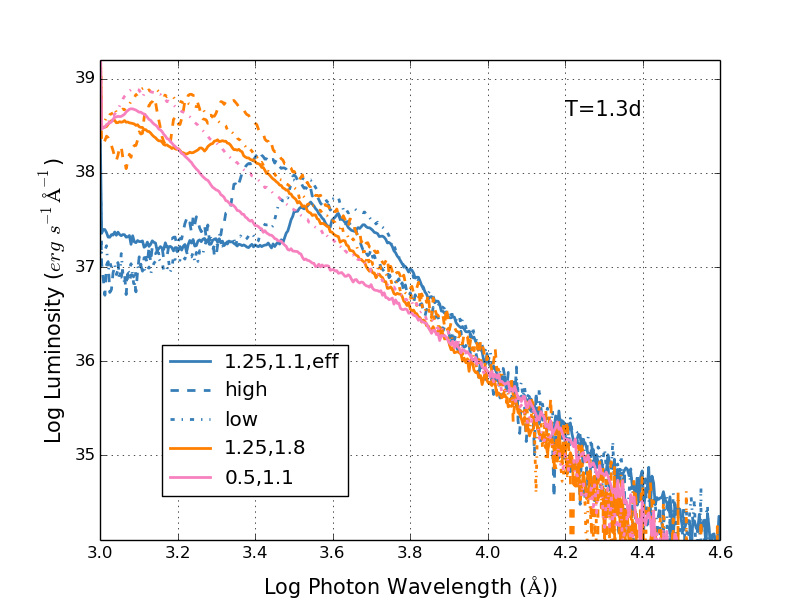}
    \includegraphics[width=0.45\textwidth]{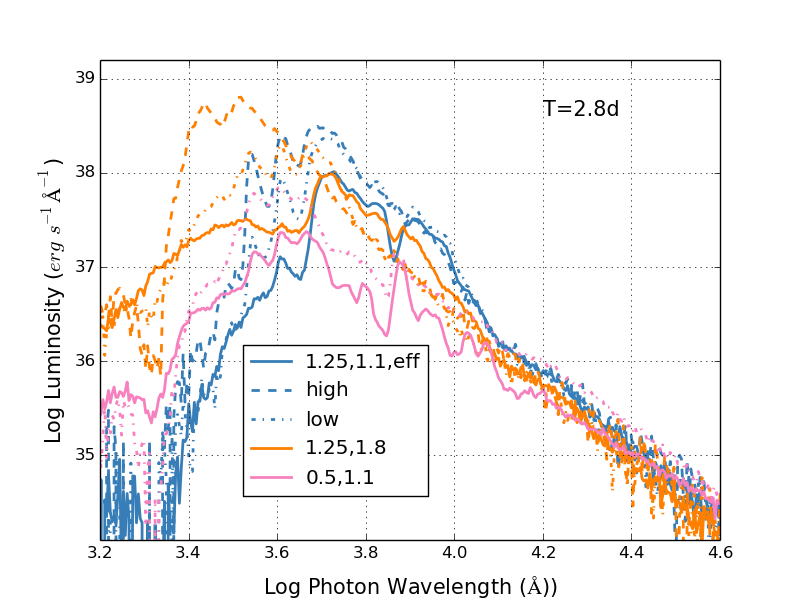}
    \includegraphics[width=0.45\textwidth]{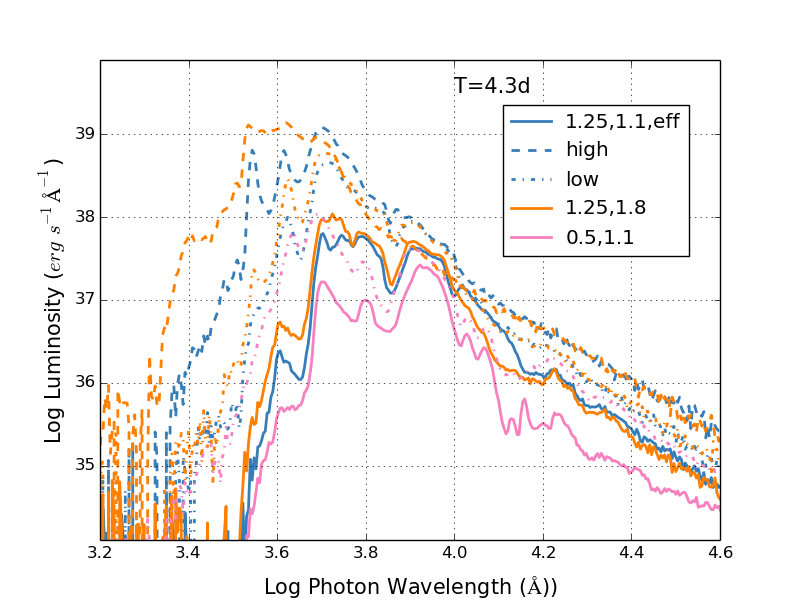}
    \caption{Spectra for a range of models varying the WD and NS mass as well as the disk model for our low ejecta-velocity models: high entropy, low entropy, and efficient wind. We follow these models at three early timescales at 1.3, 2.8, 4.3d. The scales on these plots are different from the spectra in Figure~\ref{fig:specat} because the features are redder at early times and brighter at late times.}
    \label{fig:specatv}
\end{figure}

The spectra and light-curves depend sensitively not only on the composition and ejecta mass, but also on the ejecta velocity. We have focused our study assuming the ejecta velocity is set to the escape velocity of the material within the disk. The material from the disk must be ejected above the escape velocity and this material will decelerate as it expands out of the potential well. Our standard model assumes that the ejecta ends after this deceleration with the escape velocity. Although it is standard to assume that the outmoving velocity scales with the escape velocity, the actual magnitude of the velocity is uncertain. Figure~\ref{fig:specatv} shows spectra from a set of models where we set the outmoving velocity to half of that of the escape velocity. In these models, the evolution is slower (brighter at late times) and line features are narrower. The features, especially at early times, can be redder.

In order to connect particular elements with spectral features, we extract the multigroup emissivities per element, at each space-time point in the ejecta, from the {\it SuperNu} simulations. Each element emissivity is scaled so that its integral over wavelength equals the integral of the spectrum over wavelength (i.e.,~the bolometric luminosity). The scaled emissivity per element is similar to a nebular LTE solution of the spectrum, contributed by each element. However, we note that the scaled emissivity is an imperfect indicator of spectral line expression, as high-emissivity wavelengths can be extinguished by the associated high optical depth.

\begin{figure}[h]
    \centering
    \includegraphics[width=0.45\textwidth]{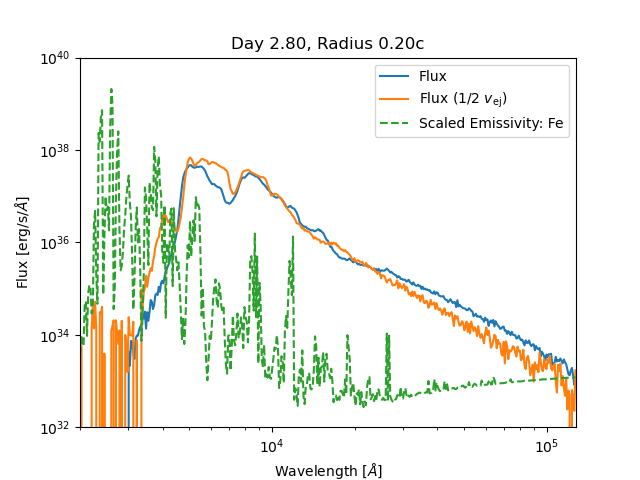}
    \includegraphics[width=0.45\textwidth]{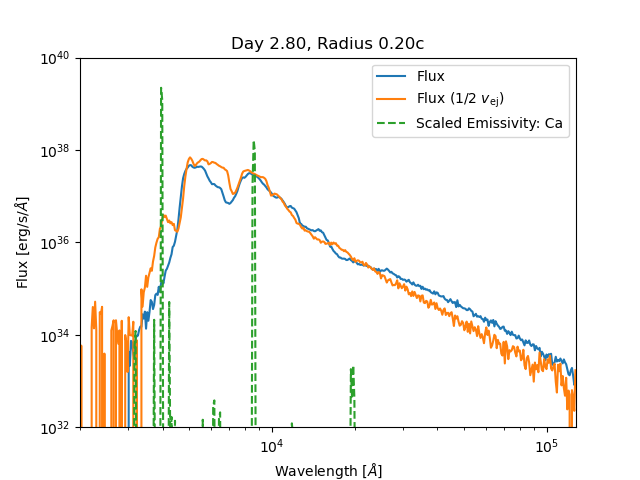} \\
    \includegraphics[width=0.45\textwidth]{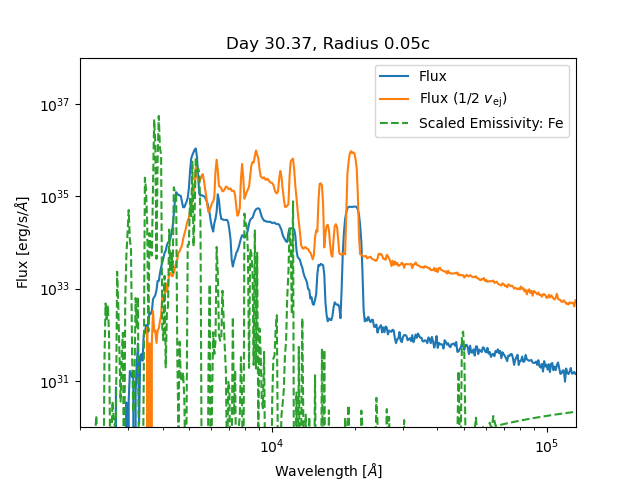}
    \includegraphics[width=0.45\textwidth]{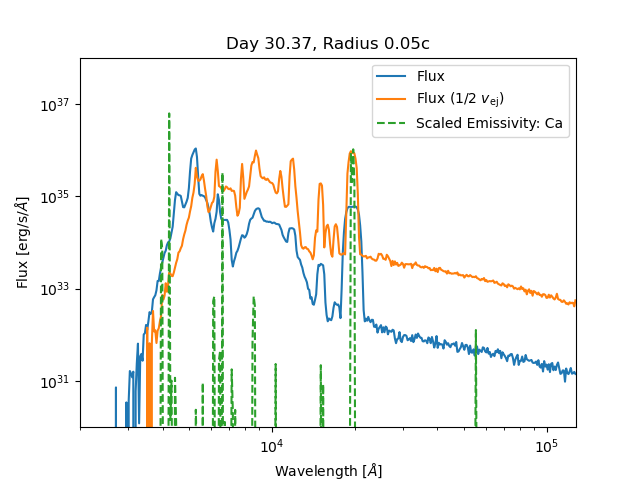}
    \caption{Spectra for a fast (blue) and slow (orange) model, along with spectrum-scaled emissivity (dashed green) of iron (left panels) and calcium (right panels) at early (top panels) and late time (bottom panels).}
    \label{fig:emiss_spec}
\end{figure}

We have examined each element's emissivity, at several ejecta velocities, and identified iron and calcium as important contributors to the spectrum. Figure~\ref{fig:emiss_spec} has spectra for a fast (blue) and slow (orange) model, along with spectrum-scaled emissivity (dashed green) of iron (left panels) and calcium (right panels) at early (top panels) and late time (bottom panels). At early time, we see that iron seems to contribute to the broad double-peak structure in the spectra, but we also note the high-emissivity region at low wavelength not expressed in the spectra; this emissivity must be suppressed by the associated high optical-depth. We also see at early time the well-known Ca II triplet line feature coincident with the second peak of the early-time spectra.

At late time, where we probe emissivity in lower-velocity ejecta layers, line features in the spectra become narrower, and we can see the iron emissivity align with several peaks in the spectra. The only element that evidently contributes to the late emission feature at $\sim$20000 $\AA$ is calcium. As in previous supernova studies with {\it SuperNu}, we have used Kurucz bound-bound data\footnote{\url{http://kurucz.harvard.edu/atoms.html}}, but this late calcium feature is also consistent with the experimental neutral Ca multiplet line observations of~\cite{humphreys1951infrared}. To our knowledge, this mid-IR calcium feature has not been discussed in the context of calcium-rich transient observations.

\subsection{Gamma Ray Bursts}

The standard Black Hole Accretion Disk gamma-ray burst model invokes the winding of magnetic-fields in a disk around a black hole~\citep{1999ApJ...518..356P}. Under this standard model, a number of progenitors and formation scenarios exist producing accretion disks around black holes that might drive the relativistic jets needed to produce these powerful explosions~\citep{Fryer_1999}. However, a black hole is not required to power a gamma-ray burst explosion. Relativistic jets have been produced in neutron star systems, either through a rapidly rotating, highly-magnetized (magnetar) neutron star or through neutron star accretion disk~\citep{2006NJPh....8..119L,2009ApJ...703..461Z,2012ApJ...761...63P,2017ApJ...839...85C,2020ApJ...901...53A,2021MNRAS.508.5390S}. 

\cite{1999ApJ...518..356P} argued that the energy derived in the disk model relied heavily on the energetics near the compact remnant in the disk. The energy that can be tapped to drive the jet is on par with its kinetic energy:
\begin{equation}
    E_{\rm disk} \approx \dot{M} v_{\rm rot}^2 = \dot{M} G M_{\rm NS}/r_{\rm NS} \approx 4 \times 10^{51} (\dot{M}/0.01\,M_\odot\,s^{-1}) {\rm \, erg \, s^{-1}}
\end{equation}
where $\dot{M}$ is the accretion rate, $v_{\rm rot}$ is the rotational velocity, $G$ is the gravitational constant, and $M_{\rm NS}=1.4\,M_\odot$ and $r_{\rm NS}=10\,{\rm km}$ are the mass and radius of the neutron star, respectively. For our accretion rates, the energy available ranges from $10^{51}-4 \times 10^{52} {\rm \, erg \, s^{-1}}$. The efficiency at which this energy can be tapped can be compared to the black hole systems studied by~\cite{1999ApJ...518..356P}. The marginally-bound radius of a rapidly-rotating, $6-7\,M_\odot$ black hole is $G\,M_{\rm BH}/c^2 \approx 10\,{\rm km}$, and we expect the efficiency at which our neutron star systems are able to extracting energy to be comparable to these black hole systems. This results in GRB luminosities in excess of $10^{49} {\rm \, erg \, s^{-1}}$, corresponding to low-luminosity GRBs (roughly an order of magnitude less powerful than standard collapsars) but with durations that can be slightly longer, similar to the GRB properties expected for BH/WD mergers~\citep{Popham1999}. Without a surrounding star to focus the jet, the jets will have wider opening angles than collapsar GRBs. We will study the distribution of these properties in more detail in a later work.

Alternatively, the jet can be driven by a highly-magnetized neutron star (magnetar) spun up by the accretion. The energy of a rotating neutron star is roughly~\citep{2019EPJA...55..132F}:
\begin{equation}
    E_{\rm rot} = 1/2 I_{\rm NS} \omega^2 = 5\times10^{50} (\omega/1000Hz)^2 \text{,}
\end{equation}
where $\omega$ is the angular velocity and $I_{\rm NS} = 10^{45} (M_{\rm NS}/M_\odot) {\rm \, cm^2 \, g}$ is the moment of inertia of a neutron star. The strong magnetic fields in a magnetar can tap this rotational energy to power a gamma-ray burst. From this, we see that we need an $\omega \approx 1000Hz$ to power a gamma-ray burst through magnetar emission. If we are using accretion to spin up the magnetar, we need to accrete an angular momentum of $I_{\rm NS} \omega(=1000Hz) = 10^{48} {\rm \, g \, cm^2 \, s^{-1}}$. Here we assume that the accreted angular momentum from our system is:
\begin{equation}
    \dot{J} = \dot{M} v_{\rm rot} r_{\rm NS} = 2.7 \times 10^{47} (\dot{M}/0.01 M_\odot s^{-1})\text{.}
\end{equation}
For our accretion rates, a few seconds of accretion is sufficient to spin up the neutron star, seemingly making this a viable mechanism to drive jets. The only difficulty with the magnetar mechanism is that this high accretion rate is believed to bury the magnetic field~\citep{2019EPJA...55..132F}, delaying the extraction of the rotational energy.

The strength of the jet is directly proportional to the accretion rate. By using the properties of the jet and the emission from the wind ejecta, we can place constraints on the component masses of the merging system. With jet and wind properties, we can potentially probe the nature of the disk wind. The high-entropy runs, which produced the strongest winds and the brightest light-curves from the ejecta, lose much of their mass well above the neutron star. Their accretion rates can be nearly an order of magnitude lower than the efficient, low-entropy wind model. The energy source for a jet produced by the high-entropy wind case will be lower by at least an order of magnitude. Understanding the nature of mass-loss in the winds is critical to understanding the strength in the jet.

We have shown that the accretion rates in these systems are sufficiently high to power a gamma-ray burst. A final complication is determining whether either of these mechanisms can produce jets with sufficiently high Lorentz factor to produce a gamma-ray burst. Even if the jet does not reach high Lorentz factors, it will affect the mass ejecta, altering the supernova light-curves discussed in Section~\ref{sec:snlc}.

\section{Conclusion\label{sec:conclusion}} 
We have presented here the accretion-disk evolution for thirteen different progenitor binaries, with three distinctly different model classes. Each model assumes a vertically-integrated, advection-dominated accretion disk, including nuclear burning, neutrino emission at high-temperature, and wind ejection from the surface of the disk. The three different classes provide different observable phenomena that may later help in validating our results. Nuclear burning in the disk produces similar results to those found in earlier NS-WD accretion-disk works. The radial composition of the disk mimics the layered structure found in massive stars. The observable nuclear yields are largely dependent on the radial strength of the wind ejecta. 

The disk models produced significant amounts of fast-moving iron-peak elements and are particularly effective at producing $^{60}$Fe when compared with other cataclysmic events. When compared with core-collapse supernovae, we find that NS-WD mergers produce and eject these heavier iron and nickel isotopes in equal or larger amounts. However, the population demographics and rates of NS-WD mergers will need to be applied to these results to determine the extent to which these mergers contribute to galactic nucleosynthetic yields. 

The observational transients expected from NS-WD mergers are incredibly diverse. The electromagnetic transients produced from our models broadly align with the recently observed AT2018kzr. The characteristics of the light-curves and spectra from our array of models lead us to conclude that this event could be a NS-WD merger. Iron and calcium are found to be the strongest contributors to our spectra, and our mid-IR Ca features should be investigated in more depth. Other observable transients, e.g., GRBs, are not ruled out if the mass of the initial disk is large enough with a low enough entropy to drive larger accretion rates onto the NS. However, the particular dynamics of the NS need to be modeled with more specificity to narrow the expected observable energies.

We found significant variation among the different classes of models. Our classes were set by varying the initial conditions of the disk and modifications to the accretion disk physics. We found the largest difference to be caused by varying the initial conditions, i.e., the entropy of the initial disk material. Further investigation and improvements to the initial composition are required to refine the results from these events. The accretion disk physics itself can continue to be improved as well. The advection-dominated assumption may not be applicable to all radii in the disk, particularly the outer regions, where we find that this assumption combines with the disk wind mechanism to strongly emit wind at large radii.

\begin{acknowledgements}

The work by MAK, CLF, RW, and WE was supported by the US Department of Energy through the Los Alamos National Laboratory. Los Alamos National Laboratory is operated by Triad National Security, LLC, for the National Nuclear Security Administration of U.S.\ Department of Energy (Contract No.\ 89233218CNA000001). Part of this work rose out of discussions at the Aspen Center for Physics, which is supported by National Science Foundation grant PHY-1607611. KB acknowledges support from the Polish National Science Center (NCN) Maestro grant (2018/30/A/ST9/00050).

\end{acknowledgements}

\bibliography{aas_paper}{}
\bibliographystyle{aas_paper}

%% the collaboration and author truncation commands are used. It has to


\end{document}